\documentstyle[12pt,epsfig]{article}
[1999/03/29 PSNFSS v.7.2
Times font as default roman
: S Rahtz]

\begin{document}
\renewcommand{\thefootnote}{\fnsymbol{footnote}}
\sloppy
\newcommand{\rp}{\right)}
\newcommand{\lp}{\left(}
\newcommand \be  {\begin{equation}}
\newcommand \bea {\begin{eqnarray}}
\newcommand \ee  {\end{equation}}
\newcommand \eea {\end{eqnarray}}

\title{Significance of log-periodic precursors to financial crashes}

\author{Didier Sornette$^{1,2}$ and Anders Johansen$^3$\\
$^1$ Institute of Geophysics and Planetary Physics \\
and Department of Earth and Space Science\\
3845 Slichter Hall, Box 951567\\
University of California, Los Angeles, California 90095-1567\\
$^2$ Laboratoire de Physique de la Mati\`{e}re Condens\'{e}e\\ CNRS UMR6622 and
Universit\'{e} de Nice-Sophia Antipolis\\ B.P. 71, Parc
Valrose, 06108 Nice Cedex 2, France\\
$^3$ The Niels Bohr Institute, University of Copenhagen\\
Blegdamsvej 17, DK-2100 Kbh. \O, Denmark
}

\maketitle

\abstract{We clarify the status of log-periodicity associated with speculative 
bubbles preceding financial crashes. In particular, we address Feigenbaum's 
[2001] criticism and show how it can be rebuked. Feigenbaum's main result is 
as follows: ``the hypothesis that the log-periodic component is present in the 
data cannot be rejected at the $95\%$ confidence level when using all the data 
prior to the 1987 crash; however, it can be rejected by removing the last year 
of data.'' ({\it e.g.}, by removing $15\%$ of the data closest to the critical 
point). We stress that it is naive to analyze a critical point phenomenon, 
{\it i.e.}, a power law divergence, reliably by removing the most important 
part of the data closest to the critical point. We also present 
the history of log-periodicity in the present context explaining its essential
features and why it may be important. We offer an extension of the rational 
expectation bubble model for general and arbitrary risk-aversion within
the general stochastic discount factor theory. We suggest
guidelines for using log-periodicity and explain how to develop and 
interpret statistical tests of log-periodicity. We discuss the issue of 
prediction based on our results and the evidence of outliers in the 
distribution of drawdowns. New statistical tests demonstrate that the 
$1\%$ to $10\%$ quantile of the largest events of the population of drawdowns 
of the Nasdaq composite index and of the Dow Jones Industrial Average index
belong to a distribution significantly different from the rest of the 
population. This suggests that very large drawdowns result from an 
amplification mechanism that may make them more predictable than smaller
market moves.}

\maketitle

\thispagestyle{empty}
\pagenumbering{arabic}
\newpage

\section{Introduction}

A market crash occurring simultaneously on the large majority of the worlds 
stock markets, as witnessed in Oct. 1987, amounts to a quasi-instantaneous 
evaporation of trillions of dollars. In present values (June. 2001), 
a worldwide stock market crash of $30\%$ indeed would correspond
to an absolute loss of about 13 trillion dollars!
Market crashes can thus swallow years of pension and
savings in an instant. Could they make us suffer even more by being the precursors or 
triggering factors of major recessions as in 1929-33 after the great
crash of Oct. 1929? Or could they lead to 
a general collapse of the financial and banking system as seems to
have being barely avoided several times in the not-so-distant past?

Stock market crashes are also fascinating because they personify the class of
phenomena known as ``extreme events''. Extreme events are characteristic of
many natural and social systems, often
refered to by scientists as ``complex systems''.

Here, we present an up-to-date synthesis of the status of the concept proposed
several years ago [Sornette et al., 1996] that market crashes
are preceded by specific log-periodic patterns developing years in advance.
Section 2 summarizes how this theory of logperiodic crash precursors
emerged. Section 3 explains that log-periodicity is associated with the 
symmetry of discrete scale invariance. Section 4 stresses the importance
of log-periodicity to help constraining the future path of stock markets.
Section 5 extends our previous formulation of the
rational expectation bubble model of stock prices
preceeding crashes to the case of arbitrary risk aversion. Section 6 
clarifies the distinction between unconditional 
(ensemble average) and conditional (specific price trajectory) returns
to explain how the bubble price and probability of crashes are intertwinned.
Section 7 presents figures summarizing the empirical evidence for log-periodicity
as well as nine new cases.  Section 8 offers a brief ``log-periodicity user's 
guideline''. Section 9 summarizes previously developed statistical tests of 
log-periodicity. Section 10 discusses crash prediction using log-periodicity,
presents a new statistical test on the Hong Kong index and provides a 
quantitative formula to assess the statistical significance of a given prediction
track record using the crash ``roulette''. Section 10 also discusses some
implications of crash prediction and presents an up-to-date assessement of
a prediction on the Nikkei index issued in Jan. 1999. Section 11
discusses the problem of charaterizing very large drawdowns (cumulative losses)
as outliers. A novel maximum likelihood testing approach is presented which
confirms the proposal that very large losses belong to a different distribution
and may result from an amplifying mechanism. Section 12 concludes.

\section{A little history along the way}

To complement the alphabetical listing of the contributors to the literature on
log-periodicity concerned with stock market crashes [Feigenbaum, 2001],
let us recall a little bit of history.

The search for log-periodicity in real data started in 1991. While working 
on the exciting challenge of predicting the failure of pressure tanks made 
of kevlar-matrix and carbon-matrix composites constituting essential elements
of the European Ariane IV and V rockets and also used in satellites for 
propulsion, one of us (DS) realized that the rupture of complex composite material 
structures could be understood as a cooperative phenomenon leading to specific 
detectable critical signatures. The power laws and associated complex exponents
and log-periodic patterns discovered in this context and published only much 
later [Anifrani et al., 1995], were found to perform quite reliably [Anifrani 
et al., 1999] for prediction purposes. The concept that rupture in 
heterogeneous material is critical has since been confirmed by a series of 
numerical [Sahimi and Arbabi, 1996; Johansen and Sornette, 1998a] and 
experimental works [Garcimartin et al., 1997; Guarino et al., 1998; 1999; 
Johansen and Sornette, 2000a; Sobolev and Tyupkin, 2000]. A prediction 
algorithm has been patented and is now used routinely on these pressure 
tanks as a standard qualifying procedure.

In 1995, the extension of this concept to earthquakes was proposed by 
Sornette and Sammis [1995]
(see also [Newman et al., 1995] who elaborated on this idea using a 
hierarchical fiber bundle model). This suggestion was consolidated on
chemical precursors to the Jan. 1995 Kobe earthquake [Johansen et 
al., 1996; 2000a] and has since been analyzed rather thoroughly
[Saleur et al., 1996; Huang et al., 2000a; 2000b;
Main et al., 2000; Sobolev and Tyupkin, 2000]. Here, the evidence is perhaps
even more controversial than in the financial context,
as earthquake data are complex, spatio-temporal and sparse.
This journal is not the place for a debate on the application to earthquakes
which is left for another audience.

At approximately the same time as the extension to earthquakes was proposed,
Feigenbaum and Freund [1996] and Sornette et al. [1996] independently suggested
that essentially the same concept should also apply to large 
financial crashes. Since then,
we and collaborators [Johansen, 1997; Sornette and Johansen, 1997; 1998a;
Johansen and Sornette, 1999a; 1999b; 1999c; 2000b; 2000c; 2001a;
Johansen et al., 1999; 2000b] and other groups [Vandewalle et al., 
1998a; 1998b; 1999;
Gluzman and Yukalov, 1998; Feigenbaum and Freund, 1998; Drozdz et al., 1999]
have reported a large number of cases as well as provided an underlying
theoretical framework based on the economic framework of rational 
expectation (RE) bubbles
(see [Sornette and Malevergne, 2001] for a recent synthesis on RE 
bubbles and Canessa [2000]
for a different theoretical route).
The yet most complete compilation of ``crashes as critical points'' has been 
published in [Johansen et al., 1999; Johansen and Sornette, 1999; 2001a].

\section{Status of log-periodicity}

First, let us stress what is not log-periodicity: it is not a 
specific signature of a critical point in the same way that scale 
invariance is not a specific property of second order phase transitions. 
True, systems at their critical point are endowed 
with the symmetry
of scale invariance. However, a system exhibiting the symmetry of 
scale invariance
is not necessarily at a critical point [Dubrulle et al., 1997]. In 
the same way,
a certain class of systems at their critical points exhibit log-periodicity.
The converse is not true: there are many systems far from criticality which
nevertheless have log-periodic behaviors (see [Sornette, 1998] for a review).

Log-periodicity is an observable signature of the symmetry of
a {\em discrete} scale invariance (DSI). DSI is a weaker symmetry than
(continuous) scale invariance. The latter is the symmetry of a system which 
manifests itself such that an observable ${\cal O}\lp x\rp $ as a function 
of the  ``control'' parameter $x$ is scale invariant under the change 
$x \to \lambda x$ for arbitrary $\lambda$, {\it i.e.}, a number  $\mu\lp 
\lambda\rp$ exists such that
\be
{\cal O} (x) = \mu\lp \lambda\rp {\cal O} (\lambda x) ~~~~.
\label{one}
\ee
The solution of (\ref{one}) is simply a power law ${\cal O}(x) = x^{\alpha}$,
with $\alpha = - {\log \mu \over \log \lambda}$, which can be verified 
directly by  insertion. In DSI, the system or the observable obeys scale 
invariance  (\ref{one}) only for {\em specific} choices of the magnification 
factor $\lambda$, which form in general an infinite but countable set of 
values $\lambda_1, \lambda_2, ...$ that can be written as $\lambda_n = 
\lambda^n$. $\lambda$ is the fundamental scaling ratio determining the 
period of the resulting log-periodicity. This property can
be qualitatively seen to encode a {\it lacunarity} of the fractal structure.
The most general solution of (\ref{one}) with $\lambda$ (and 
therefore $\mu$) is
\be \label{eqdsi}
{\cal O}(x) = x^{\alpha}~P\left({\ln x \over \ln \lambda}\right)
\ee
where $P(y)$ is an arbitrary periodic function of period $1$ in the
argument, hence the name log-periodicity.
Expanding it in Fourier
series $\sum_{n=-\infty}^\infty  c_n \exp\left(2n\pi i{\ln x\over\ln 
\lambda}\right)$,
we see that ${\cal O}(x)$ becomes a sum of power laws with the 
infinitely discrete
spectrum of complex exponents $\alpha_n = \alpha + i 2 \pi n /\ln 
\lambda$, where $n$ is an arbitrary integer. Thus, DSI leads to power laws with 
complex exponents, whose observable signature is log-periodicity. Specifically,
for financial bubbles prior to large crashes, it has been established that a first 
order representation of eq. (\ref{eqdsi})
\be \label{eqlppow}
I\lp t\rp = A + B\lp t_c -t\rp^\beta + C\lp t_c -t\rp^\beta 
\cos \lp \omega \ln \lp t_c -t\rp
-\phi \rp
\ee
captures well the behaviour of the market price $I\lp t\rp$ prior to a crash or
large correction at a time $\approx t_c$. 

There are many mechanisms known to generate log-periodicity [Sornette, 1998].
Let us stress that various dynamical mechanisms generate log-periodicity,
without relying on a pre-existing discrete hierarchical structure.
Thus, DSI may be produced dynamically (see in particular
the recent nonlinear dynamical model introduced in [Ide and Sornette, 2001;
Sornette and Ide, 2001]) and does not need to be pre-determined 
by {\it e.g.}, a geometrical network.
This is because there are many ways to break a symmetry, the subtlety here
being to break it only partially. Thus, log-periodicity per se is not 
a signature
of a critical point. Only within a well-defined model and mechanism 
can it be used as a
potential signature. If log-periodicity is present at certain times 
in financial data then, without any more information, this
only suggests that these periods of time present a kind of scale invariance
in their time evolution (we are not referring here to the power law
tails of returns [Mandelbrot, 1963; de Vries, 1994; Mantegna and Stanley, 1995;
Gopikrishnan et al., 1998], which is a completely different property).

\section{Why should we be interested in log-periodicity?}

Feigenbaum[2001] writes: ``[...] log-periodic oscillations [...] 
would obviously be of
practical value to financial economists if they could use these oscillations to
forecast an upcoming crash.'' What an understatement!

Predictions of trend-reversals or changes of regime is the focus
of most forecast efforts in essentially all domains of applications,
such as in economy, finance, weather, climate, etc. However, it is
noteworthy difficult and unreliable. It is possibly the
most difficult challenge and arguably the most interesting.
It may be useful to recall that in the 1970's there was a growing concern
among scientists and government agencies
that the earth was cooling down and might enter a new ice-age similar
to the previous little 1400-1800 ice-age
or even worse (M. Ghil, private communication
and see [Budyko, 1969; Sellers, 1969; US Committee of the 
Global Atmospheric Research Program, 1975;
Ghil and Childress, 1987])! Now that global warming is almost universally 
recognized, this indeed shows how short-sighted  predictions are! The 
situation is essentially the same nowadays: estimations of future slight 
changes of economic growth rates are rather good
but predictions of recessions and of crashes are utterly unreliable.
The almost overwhelming consensus on the reality and
magnitude of global warming is based on a clear trend over 
the 20'th century that has finally
emerged above the uncertainty level. We stress that this
consensus is not based on the prediction of
a trend reversal or a cahnge of regime. In other words, scientists 
are good at recognizing a trend
once already deeply immersed in it: we needed a century of data to extract a 
clear signal of a trend on global warming.
In contrast, the techniques presently 
available to scientists are bad at predicting most changes of regime.

In economy and finance, the situation is even worse as the people's expectation
of the future, their greediness and fear all intertwine in the construction
process of the indeterminate future. On the question of prediction, Federal 
Reserve A. Greenspan said ``Learn everything you can, collect all the data, 
crunch all the numbers before making a prediction or a financial forecast. 
Even then, accept and understand that nobody can predict the future when 
people are involved. Human behavior hasn't changed; people are unpredictable. 
If you're wrong, correct your mistake and move on.'' The fuzziness resulting 
from the role played by the expectation and discount of the future on present 
investor decisions is captured by another famous quote from A. Greenspan
before the Senate Banking Committee, June 20, 1995: ``If I say something which 
you understand fully in this regard, I probably made a mistake.'' In this 
perspective, our suggestion that financial time series might contain
useful information of changes of regimes, reflecting a possible collective 
emergent behavior deserve careful attention for its numerous implications 
and applications.

There is another related reason why log-periodicity might be important.
The fundamental valuation theory provides a rational expectation estimation
of the price of an asset as the present discounted value of all future incomes
derived from this asset. However, one of the most enduring problem in 
finance is that
observed prices seem to deviate significantly and over extended time intervals
from fundamental prices. To address this problem, Blanchard [1979] and
Blanchard and Watson [1982]
introduced the concept of rational expectation (RE) bubbles,
which allow for arbitrary deviations from fundamental prices while
keeping the fundamental anchor point of economic modeling.
An important practical problem is to detect such bubbles, if they exist,
as they probably constitute one of the most
important empirical fact to explain and predict
their financial impacts (potential losses of up to
trillions of dollars during crashes and recession following these bubbles).
A large literature has indeed emerged on theoretical refinements of 
the original concept
and on the empirical detectability of RE bubbles in financial data (see
[Camerer, 1989] and [Adam and Szafarz, 1992] for a survey). Empirical
research has largely been concentrated on testing for explosive trends in the 
time series of asset prices and foreign exchange rates [Evans, 1991; Woo, 
1987], however with limited success. The first reason lies
in the absence of a general definition, as bubbles are model specific 
and generally
defined from a rather restrictive framework.  The concept of a fundamental
price reference does not necessarily exist, nor is it necessarily 
unique. Many RE
bubbles exhibit shapes that are hard to reconcile with the economic intuition
or facts [Lux and Sornette, 2001].
A major problem is that the apparent evidence for
bubbles can be re-interpreted in terms of market fundamentals that are
unobserved by the researcher. Another suggestion is that, if
stock prices are not more explosive than dividends, then it can be
concluded that rational bubbles are not present, since bubbles are
taken to generate an explosive component to stock prices.  However,
periodically collapsing bubbles are not detectable by using
standard tests to determine whether stock prices are more explosive or
less stationary than dividends [Evans, 1991]. In sum, the present evidence
for speculative bubbles is fuzzy and unresolved at best, according to the
standard economic and econometric literature. 

The suggestion that log-periodicity may be associated with bubbles 
would thus provide
a tool for their characterization and detection. In other words, we
offer the suggestion that the conundrum of bubble definition and detection
could be resolved by using the log-periodic power law structures
as one of the qualifying signatures.

\section{The rational expectation bubble model with risk aversion}

In his section 2,
Feigenbaum [2001] summarizes accurately our rational expectation bubble theory
[Johansen and Sornette, 1999a; Johansen et al., 1999; 2000b]. 
However, he suggests
that it might not be robust within a general formulation of risk aversion.
We now show that it is actually robust by solving the corresponding 
no-arbitrage
condition within the general stochastic pricing kernel framework
[Cochrane, 2001] used by Feigenbaum [2001].

We already gave [Johansen et al., 1999] two ways
of incorporating risk aversion into our model. The first one consists
in introducing a risk premium rate $r \in (0,1]$ such that the 
no-arbitrage condition
on the bubble price reads
\begin{equation}
(1-rdt) {\rm E}_t[p(t+dt)]= p(t)~,  \label{hgkak}
\end{equation}
where ${\rm E}_t[y]$ denotes the expectation of $y$ conditioned on 
the whole past
history up to time $t$.
Let us write the bubble price equation as (equation (7) in [Feigenbaum, 2001])
\be
{dp \over p} = \mu(t) dt + \sigma(t) dW - \kappa dj  ~,
\label{ghgnaql}
\ee
where $dW$ is the increment of a random walk with no drift and unit variance,
and $dj$ is the jump process equal to $0$ in absence of a crash and 
equal to $1$
when a crash occurs. The crash hazard rate $h(t)$ is given by ${\rm 
E}_t[dj]=h(t) dt$ by
definition. The condition (\ref{hgkak}) gives $\mu -r =\kappa h(t)$. The
introduction of the risk aversion rate $r$ has only the effect of 
multiplying the
price by a constant factor and does not change the results.

Another way to incorporate risk aversion is to say that
the probability of a crash in the next instant is perceived by traders
as being $K$ times bigger than it objectively is. This amounts to
multiplying the crash hazard rate $h(t)$ by $K$ and therefore does not
either modify the structure of $h(t)$.

The coefficients $r$ and $K$ both represent general aversion
of fixed magnitude against risks.  Risk aversion is a central feature
of economic theory and is generally thought to be stable within a
reasonable range being associated with slow-moving secular trends like
changes in education, social structures and technology.
Risk perceptions are however constantly changing in the course of real-life 
bubbles. This is indeed
captures by our model in which risk
perceptions quantified by $h(t)$ do oscillate dramatically throughout 
the bubble, even
though subjective aversion to risk remains stable, simply because the
{\em objective degree of risk that the bubble may burst} goes through
wild swings.

The theory of stochastic pricing kernel [Cochrane, 2001] provides
a unified framework for pricing consistently all assets under the premise that
price equals expected discounted payoff, while capturing the macro-economic 
risks underlying each security's value. This theory basically amounts to 
postulating the existence of a stochastic discount factor (SDF) $M$ that 
allows one to price all assets. The SDF is also termed the pricing kernel, 
the pricing operator, or the state price density and can be thought of as 
the nominal, inter-temporal, marginal rate of substitution for consumption
of a representative agent in an exchange economy.  Under an adequate 
definition of the space of admissible trading strategies, the product of 
the SDF with the value process of any admissible self-financing trading
strategy implemented by trading on financial, $p$, must be a martingale:
\be
M(t) p(t)= {\rm E}_t \left[p(t') M(t') \right]~, \label{martcond}
\ee
(where $t'$ refers to a future date), for no arbitrage opportunities to exist. 
In the notation of Feigenbaum [2001], $\rho(t')= M(t')/M(t)$. Expression
(\ref{martcond}) is much more general than (\ref{hgkak}) and recovers it
for $M(t) = e^{rt}$ corresponding to a constant discount rate.
The no-arbitrage condition (\ref{martcond}) expresses that $p M$ is a
martingale for any admissible price $p$. Technically, this amounts to 
imposing that the drift of $p M$ is zero.

To make further progress,
we assume the following general dynamics for the SDF
\be
\frac{dM(t)}{M(t)}  = -r(t)~ dt  - \phi(t)~dW(t) - g(t) d{\hat W}~. \label{sdf}
\ee
The drift $-r(t)$ of $M$ is justified by the well-known martingale 
condition on the product of the bank account 
and the SDF [Cochrane, 2001]. Intuitively, it retrieves the usual simple form
of exponential discount with time. 
The process $\phi$ denotes the market price of risk, as measured by the 
covariance of asset returns with the SDF. Stated differently, $\phi$ is
the excess return over the spot interest rate that assets must earn per unit of
covariance with $W$. The last term $- g(t) d{\hat W}$ embodies all other 
stochastic factors acting on the SDF, which are orthogonal to the stochastic 
process $dW$. Writing that the process $p M$ is a martingale and applying
Ito's calculus gives the following equation
\be
\mu(t) - r(t) = \kappa h(t) + \sigma(t) \phi(t)~,
\label{gblanvla}
\ee
where we have used that ${\rm E}_t [d{\hat W}]=0={\rm E}_t [dW \cdot 
d{\hat W}]$.
The case $r(t)=\phi(t)=0$ recovers our previous formulation (equation (9) of
[Feigenbaum, 2001]). The expected price conditioned on no crash 
occurring ($dj=0$) is obtained
by integrating (\ref{ghgnaql}) with (\ref{gblanvla})
\be
{\rm E}_{t_0}[p(t)] = p(t_0) L(t) ~ \exp \left( \kappa \int_{t_0}^t 
d\tau  h(\tau) \right)~,
\label{fjjma}
\ee
where
\be
L(t) = \exp \left(\int_{t_0}^t d\tau  \left[ r(\tau) + \sigma(\tau) 
\phi(\tau) \right] \right)~.
\ee
For $r(t)=\phi(t)=0$, $L(t)=1$, we recover our previous result
[Johansen and Sornette, 1999a; Johansen et al., 1999; 2000b].
Expression (\ref{fjjma}) shows how the possible log-periodic 
structures carried by the crash hazard rate $h(t)$ can be distorted by 
the SDF. It is worth noting that the log-periodic signal
$\kappa \int_{t_0}^t d\tau  h(\tau)$, if present, does not disappear.
Only a possibly noisy signal $\ln L(t)$ is added to it. The problem is then to
detect a regular (log-periodic) oscillation embedded within a noisy 
signal. For this,
we have proposed the powerful Lomb spectral analysis in the variable
$\ln t_c-t$ [Johansen et al., 1999], where $t_c$
is the critical time corresponding to the theoretical end of the bubble.

We thus see that the most general form of risk aversion does not invalidate our
theory. We note in passing that the SDF is not different from one 
agent to the next, as argued by Feigenbaum [2001], because
it described the aggregate perception by the rational agents of the 
level of risks [Cochrane, 2001].

\section{Average returns versus conditional returns}

In several instances in his article, Feigenbaum stresses that our model
disallows any positive profit, because we use the no-arbitrage condition.
The same error has been made previously by Ilinsky [1999].

Their argument is basically that the martingale condition (\ref{martcond})
leads to a model which ``assumes a zero return as the best prediction for
the market.'' Ilinsky in particular continues with
``No need to say that this is not what one expects from
a perfect model of market bubble! Buying shares, traders expect the price
to rise and it is reflected (or caused) by their prediction model. 
They support the bubble and the bubble supports them!''.
In other words, Feigenbaum and Ilinsky criticize a key economic 
hypothesis of our model:
market rationality.

This misunderstanding addresses a rather subtle point of the model and
stems from the difference between two different types of returns\,:
\begin{enumerate}
\item the unconditional return is indeed zero as seen from (\ref{hgkak}) or
(\ref{martcond}) and reflects the fair game condition.

\item The conditional return $\mu(t)$, conditioned upon no crash 
occurring between time $t$ and time $t'$, is non-zero and is given by 
(\ref{gblanvla}). If the crash hazard rate is increasing with time,
the conditional return will be accelerating precisely because the crash
becomes more probable and the investors must be remunerated for their
higher risk in order to participate in the continuation of the bubble.
\end{enumerate}

The vanishing unconditional average return
takes into account the probability
that the market {\em may} crash. Therefore, {\em conditionally}
on staying in the bubble (no crash yet), the market must rationally
rise to compensate buyers for having taken the risk that the market {\em
could} have crashed.

Our model thus allows (conditional) profits. Only when summing over
many bubbles and many crashes will these profits average out to zero. 
As a consequence,
it is possible to construct trading strategies to test for the existence of
abnormal profit. This has been done with positive results
(Lin., P. and Ledoit. O., 1999, private communication).

\section{Log-periodic bubbles prior to crashes}

\begin{figure}
\begin{center}
\epsfig{file=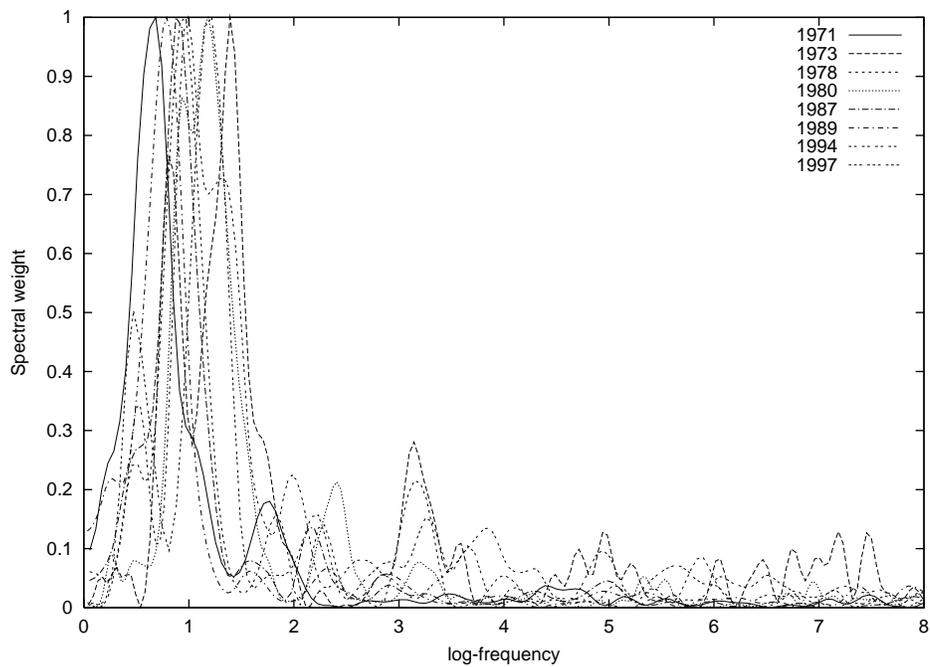}
\caption{\protect\label{lomballhk} The Lomb periodogram of the log-periodic
component of the Hang-Seng price index (Hong Kong) for the 8 bubbles 
followed by crashes observed in figure \protect\ref{scanhk}, ending in Oct. 
1971, in Feb, 1973, in Sept. 1978, in Oct. 1980, in Oct. 1987, in April 1989,
in Jan. 1994 and in Oct. 1997. See [Johansen et al., 1999] for 
details on how to calculate the Lomb periodogram.
}
\end{center}
\end{figure}

\begin{figure}
\begin{center}
\epsfig{file=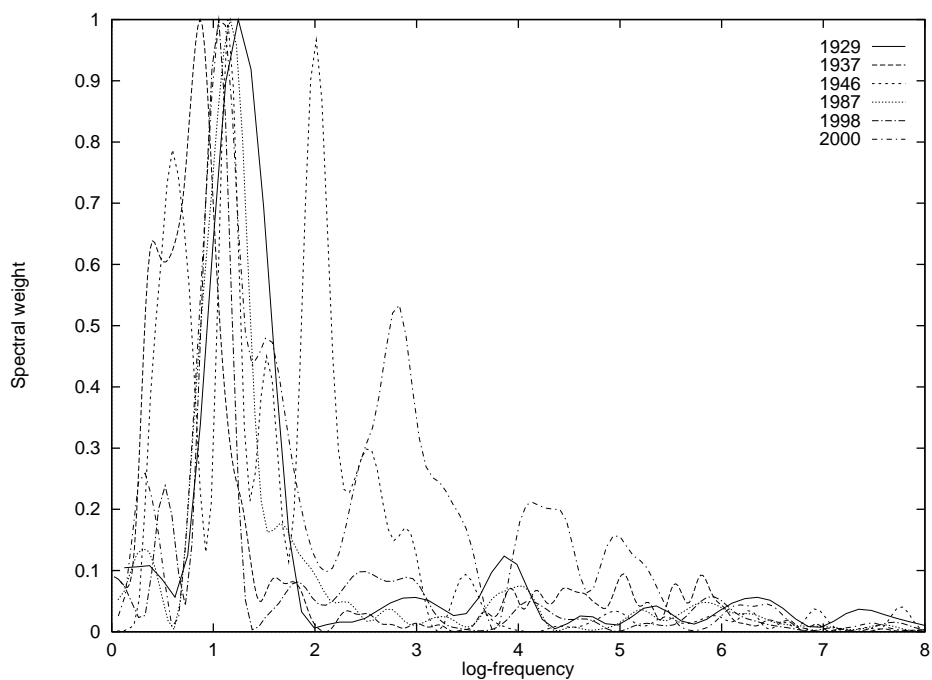}
\caption{\protect\label{lomballws} Log-periodic spectra for all the major 
bubbles ending in crashes on the Dow Jones and SP500 index in the twentieth century 
as well as the Nasdaq crash of 2000. Observe that the sub-harmonics (half 
log-frequency) and two harmonics $2f$ and $3f$ are quite strong in some of 
the data sets. See [Johansen et al., 1999] for details on how to calculate 
the Lomb periodogram.
}
\end{center}
\end{figure}

\begin{figure}
\begin{center}
\epsfig{file=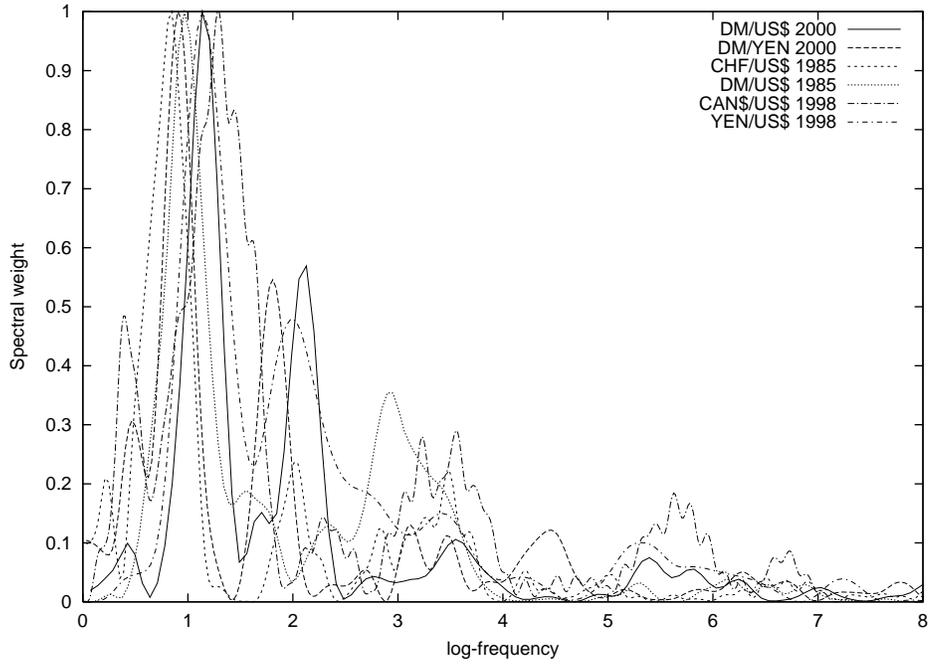}
\caption{\protect\label{lomballfx} Log-periodic spectra for the major recent
bubbles on currencies.  See [Johansen et al., 1999] for details on how to 
calculate the Lomb periodogram.
}
\end{center}
\end{figure}

In figures \ref{lomballhk} to \ref{lomballfx}, we show a spectral analysis 
of the log-periodic component of all previously identified and published 
log-periodic bubbles prior to crashes on the major financial markets as well 
as nine new cases for completeness\footnote{The Yen versus DM mini-crash
was in fact presented by one of us (AJ) at a talk on the Niels Bohr Inst. 
on 3. Jan. 2000, {\it i.e.}, prior to the actual date of the mini-crash. Thus
the talk formally constitutes a prediction of the mini-crash. 
The evidence on the two crashes on Wall Street in 1937 and 1946 was presented
at a talk at the investment company PIMCO, Newport Beach in May 2000.}.

These new cases consist in 
\begin{enumerate}
\item two crashes on currencies, the Yen versus Euro (expressed in Deutsch mark)
mini-crash with a drop of more than $7\%$ on Jan. 2000 and the US\$ versus Euro event 
(expressed in Deutsch mark) with a drop of more than $13\%$ on Oct. 2000;

\item two crashes of the DJIA and SP500 on the US market in early 1937 and in mid-1946;

\item five crashes of the Heng Seng index of the Hong Kong market, of
April 1989, Nov. 1980, Sept. 1978, Feb. 1973 and Oct. 1971, which complement
the three crashes on Oct. 1987, Jan. 1994 and Oct. 1997 previously reported
in [Johansen and Sornette 2001a]. The positions of these five crashes
are pointed out in figure \ref{scanhk} which presents the 
Hang-Seng composite index from Nov. 1969 to Sept. 1999. 
\end{enumerate}

\begin{figure}
\begin{center}
\epsfig{file=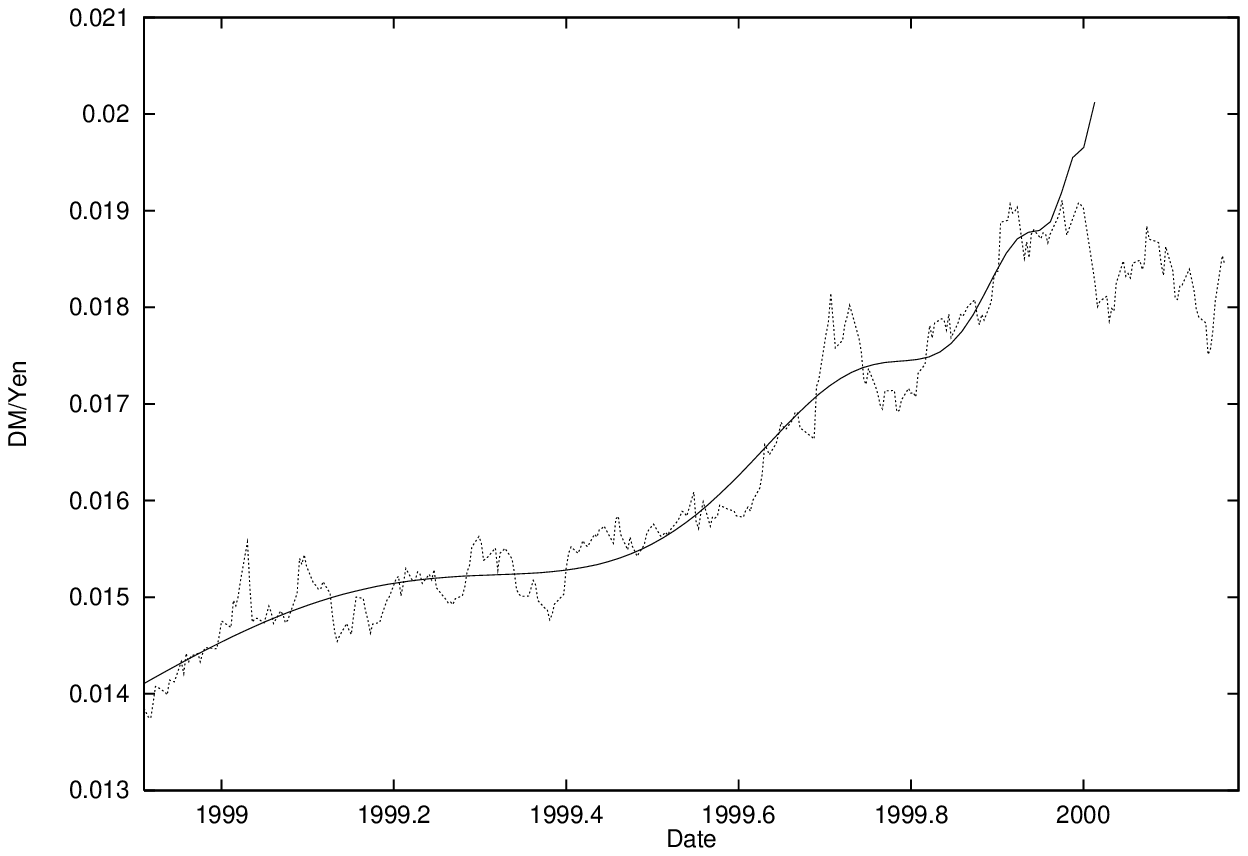}
\caption{\label{yendm2000} Yen versus Euro (expressed in DM)
 mini-crash (drop $>7\%$) of Jan 2000. The 
parameter values of the fit with eq. \protect\ref{eqlppow} are $A \approx 
0.021$, $B\approx -0.0067$, $C\approx 0.00048$, $\beta \approx 0.45$, $t_c 
\approx 2000.0$, $\phi \approx  -0.5$ and $\omega \approx  5.7$.
}
\end{center}
\end{figure}

\begin{figure}
\begin{center}
\epsfig{file=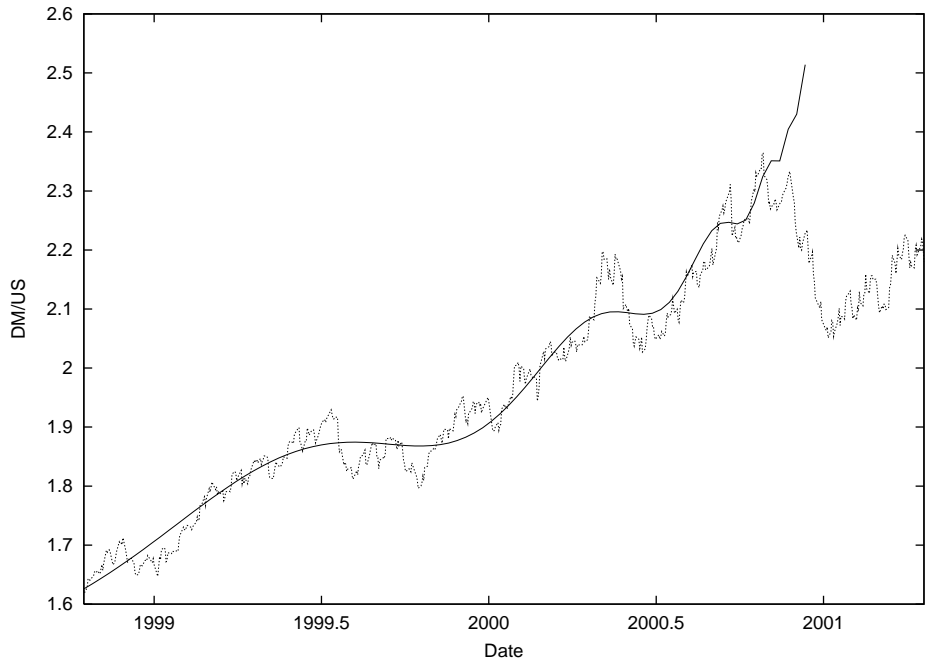}
\caption{\label{usdm2000} US\$ versus Euro (expressed in DM) crash (drop $>13\%$) of 2000. 
The parameter
values of the fit with eq. \protect\ref{eqlppow} are $A \approx 2.6$, $B\approx
-0.65$, $C\approx -0.05$, $\beta \approx 0.44$, $t_c \approx 2000.88$, 
$\phi \approx  0.0$ and $\omega \approx  7.3$.
}
\end{center}
\end{figure}

\begin{figure}
\begin{center}
\epsfig{file=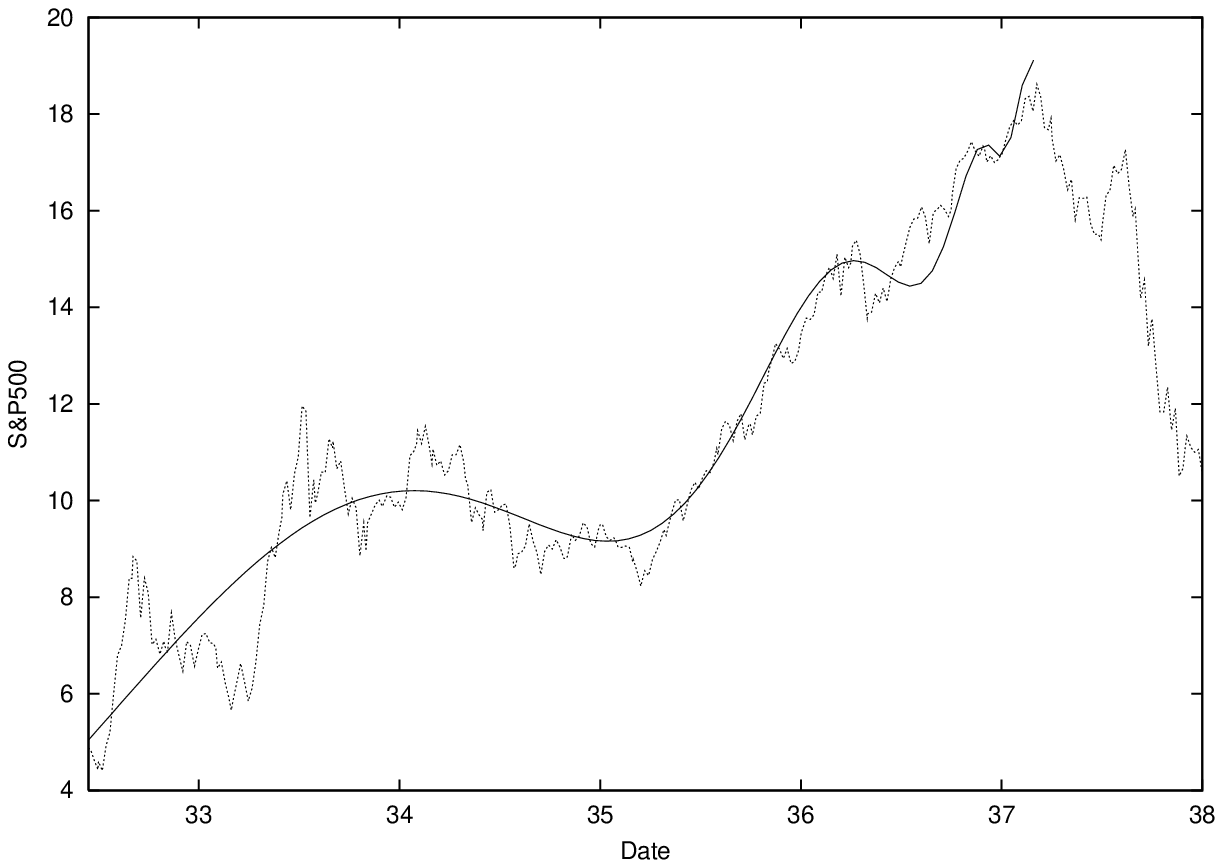}
\caption{\label{ws37} Wall Street crash of 1937. The parameter values of the 
fit with eq. \protect\ref{eqlppow} 
are $A \approx 19.9$, $B\approx -6.1$, $C\approx 1.1$, $\beta \approx 
0.56$, $t_c \approx 1937.19$, $\phi \approx 0.1$ and $\omega \approx  5.2$.
}
\end{center}
\end{figure}

\begin{figure}
\begin{center}
\epsfig{file=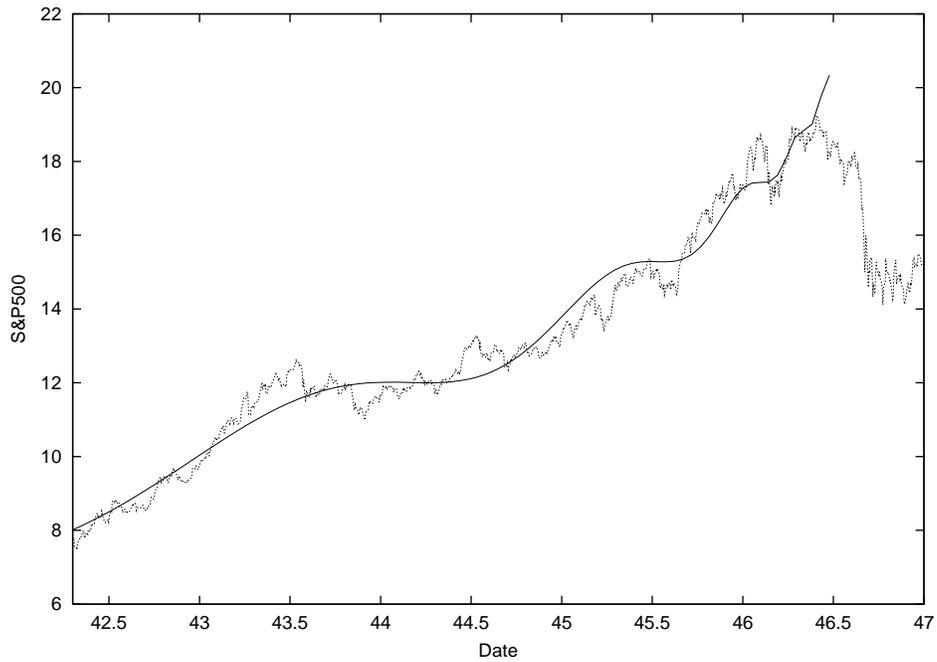}
\caption{\label{ws47} Wall Street crash of 1946. The parameter values of the 
fit with eq. \protect\ref{eqlppow}
are $A \approx 21.5$, $B\approx -6.3$, $C\approx -0.44$, $\beta \approx
0.49$, $t_c \approx 1946.51$, $\phi \approx  1.4$ and $\omega \approx  7.2$.
}
\end{center}
\end{figure}

\begin{figure}
\begin{center}
\epsfig{file=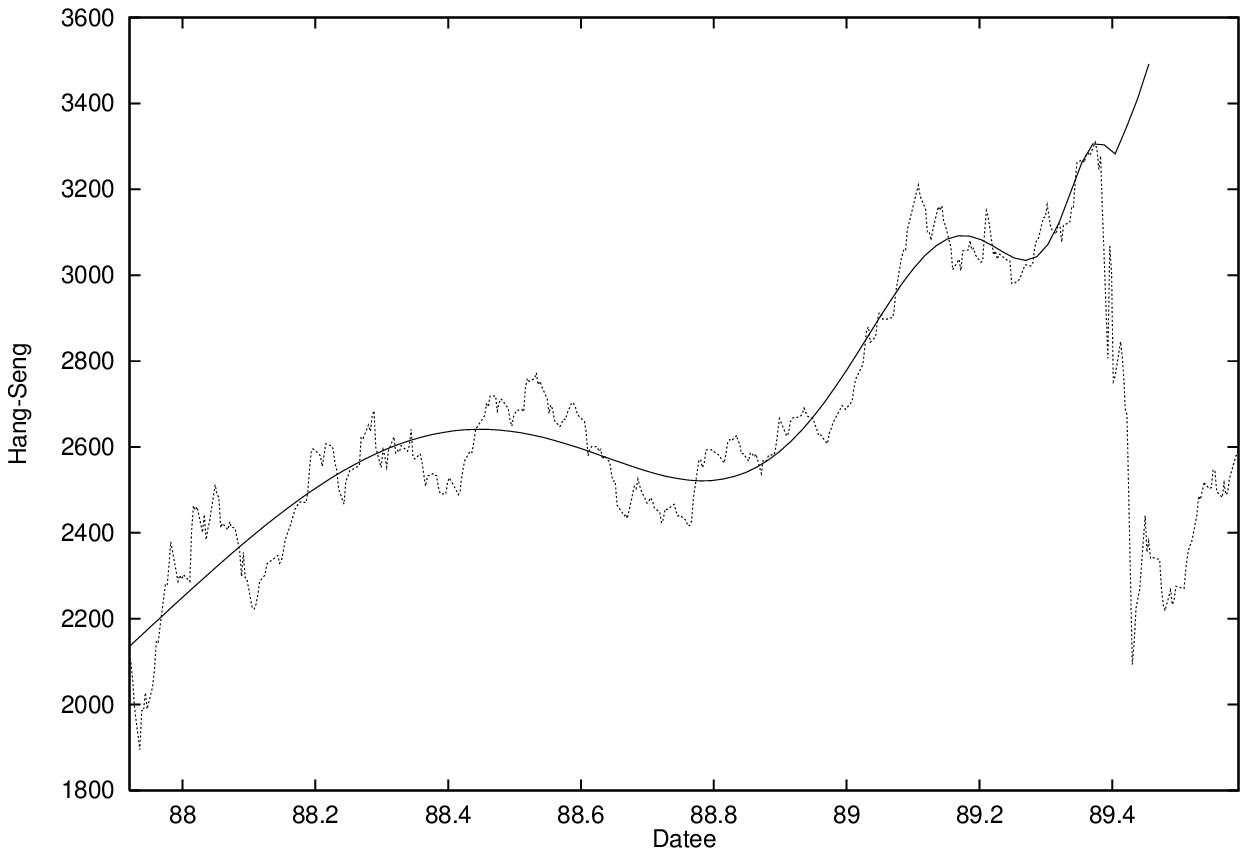}
\caption{\label{hk89} Hong-Kong crash of 1989. The parameter values of the
fit with eq. \protect\ref{eqlppow} are $A \approx 3515$, $B\approx -1072$, 
$C\approx 225$, $\beta \approx 0.57$, $t_c \approx 1989.46$, $\phi \approx  
0.5$ and $\omega \approx  4.9$.
}
\end{center}
\end{figure}

\begin{figure}
\begin{center}
\epsfig{file=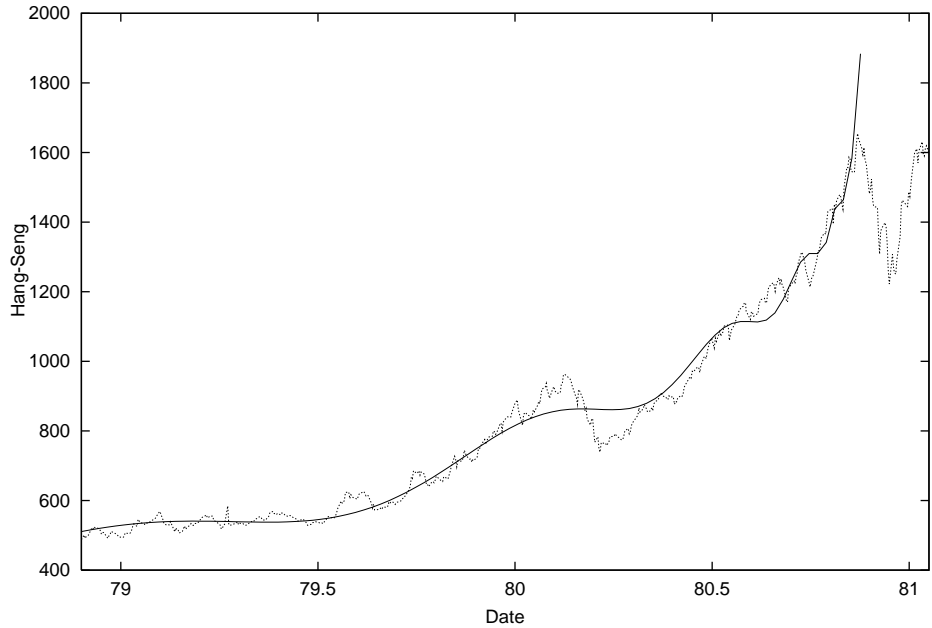}
\caption{\label{hk80} Hong-Kong crash of 1980. The parameter values of the
fit with eq. \protect\ref{eqlppow} are $A \approx 2006$, $B\approx -1286$, 
$C\approx -55.5$, $\beta \approx
0.29$, $t_c \approx 1980.88$, $\phi \approx  1.8$ and $\omega \approx  7.2$.
}
\end{center}
\end{figure}

\begin{figure}
\begin{center}
\epsfig{file=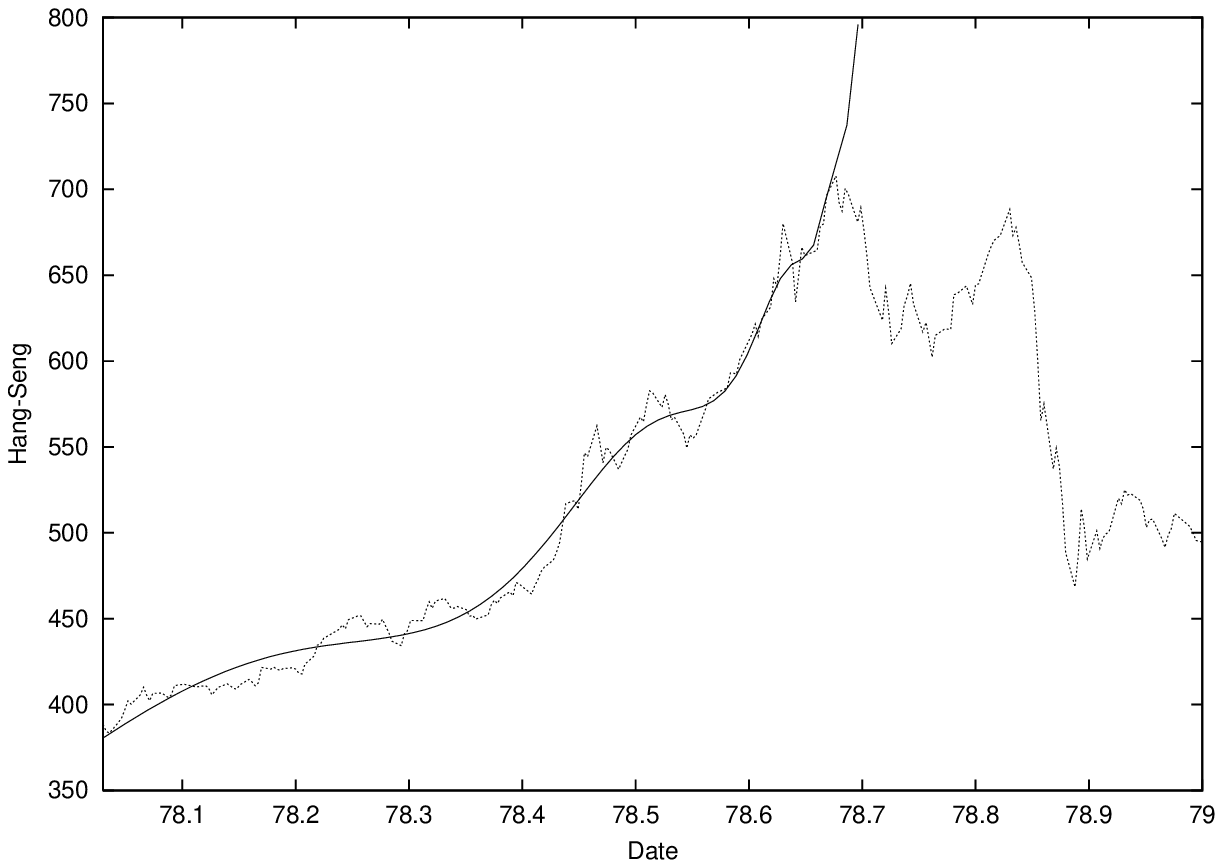}
\caption{\label{hk78} Hong-Kong crash of 1978. The parameter values of the fit 
with eq. \protect\ref{eqlppow} are $A \approx 824$, $B\approx -538$, $C\approx 
-28.0$, $\beta \approx 0.40$, $t_c \approx 1978.69$, $\phi \approx  -0.17$ and 
$\omega \approx 5.9$.
}
\end{center}
\end{figure}

\begin{figure}
\begin{center}
\epsfig{file=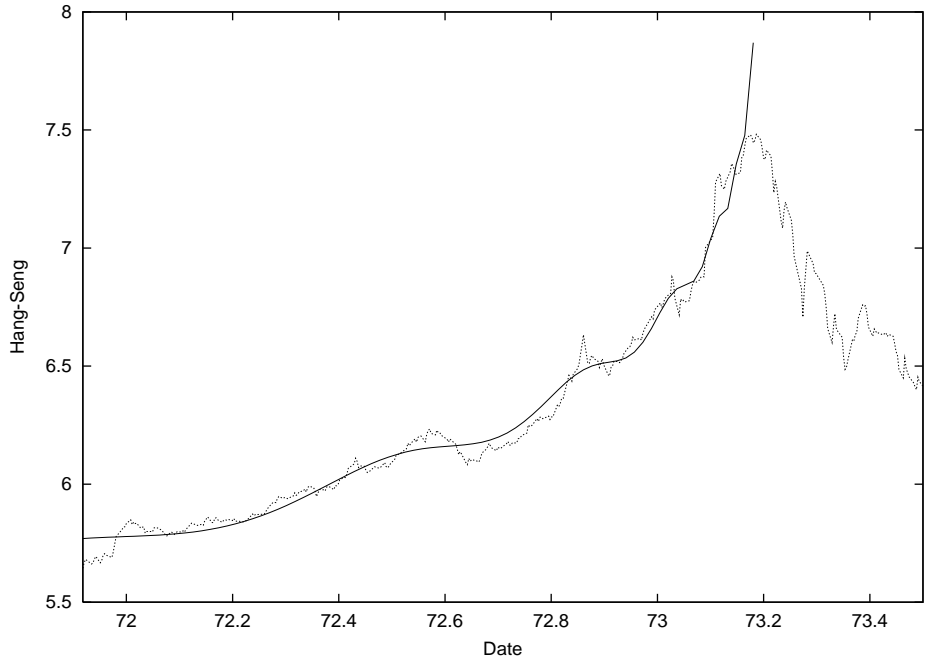}
\caption{\label{hk73} Hong-Kong crash of 1973. The parameter values of 
the fit with eq. \protect\ref{eqlppow} are $A \approx 10.8$, $B\approx -5.0$, 
$C\approx -0.05$, $\beta \approx 0.11$, $t_c \approx 1973.19$, $\phi \approx  
-0.05$ and $\omega \approx  8.7$. Note that for this bubble, it is the 
logarithm of the index which is used in the fit.
}
\end{center}
\end{figure}

\begin{figure}
\begin{center}
\epsfig{file=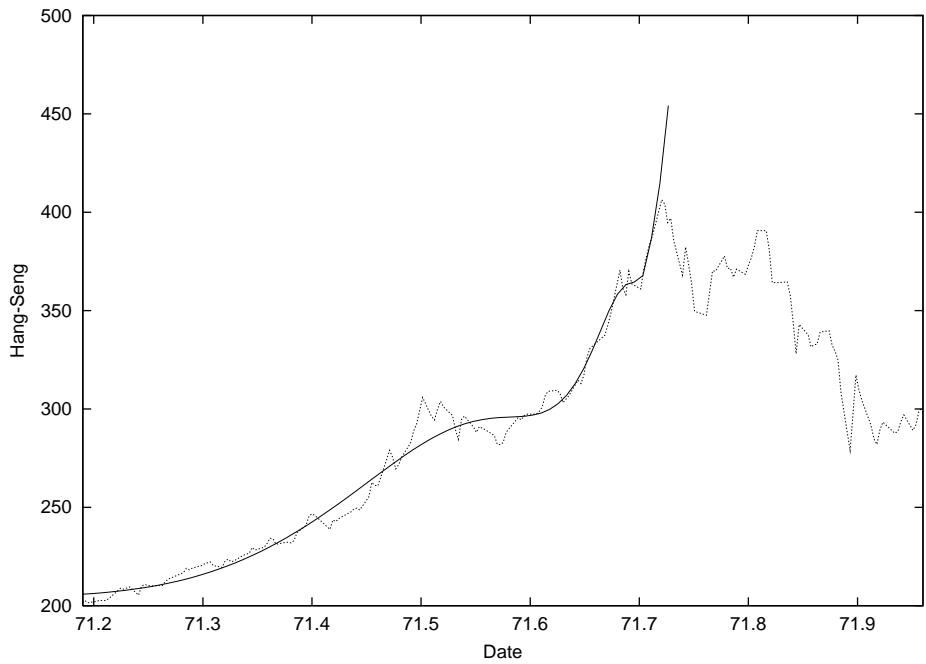}
\caption{\label{hk71} Hong-Kong crash of 1971. The parameter values of the 
fit with eq. \protect\ref{eqlppow} are $A \approx 569$, $B\approx -340$, 
$C\approx 17$, $\beta \approx 0.20$, $t_c \approx 1971.73$, $\phi \approx  
-0.5$ and $\omega \approx  4.3$.
}
\end{center}
\end{figure}

The fits with the log-periodic formula (\ref{eqlppow})
to these nine new cases are shown in figures \ref{yendm2000} to 
\ref{hk71}. Five of them are the first four bubbles on the Hang-Seng index 
ending respectively in Oct. 1971, in Feb, 1973, in Sept. 1978 and in Oct. 1980
plus one additional event ending in April 1989.
Among these five events of the Hong Kong market, the first 
(Oct. 1971), third (Sept. 1978)  and fifth (April 1989) are significantly
weaker than all the others and occur over time periods shorter than most
of the previous ones, hence we expect the recovery of the log-periodic
parameters, and in particular of the angular log-frequency $\omega$,
to be less precise.

Consistently, all twenty cases ($8$ crashes on the sole Hong Kong market,
$6$ crashes on the major stock markets and $6$ crashes on currencies)
 point to a log-periodic component in the 
evolution of the market price with a log-frequency $\omega /2\pi \approx 1.1 
\pm 0.2$ except for the two first Hong-Kong cases of 1971 and 1973 and the
crash of April 1989 where the 
value for $\omega$ is slightly lower/higher. Furthermore, as can be seen in 
the captions of figures \ref{yendm2000} to \ref{hk80} for the nine new cases 
and in [Johansen et al., 1999; Johansen and Sornette, 1999; 2001a] for the 
previously published cases,  the value of the exponent $\beta$ is remarkably 
consistent found around $\approx 0.35 \pm 0.15$, however with larger 
fluctuations than for $\omega$. Of the twenty large crashes in major financial 
markets of the 20th century, seventeen occurred in the past $\approx 30$ years,
suggesting an increasing strength of collective and herding behavior over time.
That these twenty large crashes show such consistent values for the two 
meaningful parameters $\beta$ and especially $\omega$ as well as the timing 
$t_c$ of the crash is needless to say {\it quite remarkable}. The reader is 
reminded that many more cases, such as the Russian bubble followed by a crash 
in 1997 which was erronously postulated by Ilinski [Ilinski, K. 1999] not to 
contain any 
log-periodic component, and on several emergent markets have been identified 
[Johansen et al., 1999; Johansen and Sornette, 1999; 2001a].

\section{Log-periodicity user's guidelines}
 
We are obviously not going to divulge our technique and methodology for crash
prediction but instead 
offer a few common sense guidelines
to avoid the rather obvious traps in which Feigenbaum [2001] has fallen.

The first important observation is the rather large degeneracy of the 
search landscape
of the fit of the data by equation (22) in [Feigenbaum, 2001]. 
Feigenbaum presents
the two best solutions to the fit of the logarithm of the price
of the SP500 index prior to the Oct. 1987 crash
in his figures 3 and 4. These two solutions tell quite different stories.
We just note here that we would have rejected the solution of his figure 
4 as the angular
log-frequency $\omega = 2.24$ is very small, corresponding to
a huge preferred scaling ratio $\lambda = 16.5$. In addition, the 
exponent $\beta$
is larger than $1$, which as recalled correctly by Feigenbaum is
a calibrating constraint. In such complex search problems, the 
root-mean-square and $R^2$
statistics are not very discriminating and should not be taken as the 
first filters,
as already explained in our previous publications. An obvious technical reason
is that the errors or fluctuations around the log-periodic component 
are not Gaussian and instead exhibit fat tails, reflecting those of the 
distribution function of returns 
[Mandelbrot, 1963; de Vries, 1994; Mantegna and Stanley, 1995; 
Gopikrishnan et al., 1998].
We also disagree with Feigenbaum when he writes ``we are not 
consistently estimating
physical parameters''. Indeed, $\beta$ and especially $\omega$ carries a lot 
of relevant information as shown by the Lomb peridograms the 
log-periodic oscillations in [Johansen et al 1999].

Among the five examples presented in table 1 of [Feigenbaum, 2001], three
(1987, 1997\footnote{A trading strategy using put options was devised as an 
experimental test of the theory. Based on the prediction of the stock market 
turmoil at the end of October 1997, a 400 \% profit has been obtained in a two 
week period covering the mini-crash of Oct. 27 1997. The proof is available 
from a Merill Lynch client cash account release in Nov. 1997. See also the work
by A.Minguet {\it et al.~} in H. Dupois {\em Un krach avant novembre}, Tendance
18, page 26 Sept. 1997.} and 1998) are consistent with our and other authors' 
previous results. The 1974 case could be called an ``anti-bubble'' ending in a 
large positive correction or drawup, an ``anti-crash''. We have observed 
several other similar anti-bubbles [Johansen and Sornette, 1999c], some ending 
in such an ``anti-crash'' (this will be reported elsewhere). The case that
Feigenbaum makes with the 1985 event for suggesting the absence of any
discriminating power of log-periodicity is easily dismissed: we would 
never qualify it since its exponent $\beta = 3.53$ is much too large. We 
stress that it is an all too common behavior to dismiss lightly a serious
hypothesis by not taking the trouble to learn the relevant skills necessary to 
test it rigorously, as for example in [Laloux et al., 1999].

\section{How to develop and interpret statistical tests of log-periodicity}

With respect to the possible selection bias of fitted time intervals,
we stress that we already addressed specifically this question 
[Johansen et al., 2000b]. For completeness and courtesy to the reader, we 
summarize our results. We picked at random fifty $400$-week intervals in 
the period 1910 to 1996 of the logarithm of the Dow Jones Industrial Average 
and launched the same fitting procedure as
done for the 1987 and 1929 crashes. The end-date of
the 50 data sets are given in [Johansen et al., 2000b]. Of
the $11$ fits with a root-mean-square comparable with that of the two crashes,
only $6$ data sets produced values for $\beta$, $\omega$ and $\Delta_t$
which were in the same range as the values obtained for the $2$ crashes,
specifically $0.45 < \beta < 0.85$, $4 < \omega < 14$ ($1.6 < \lambda <4.8$)
and $3 < \Delta_t < 16$. All $6$ fits belonged to the periods prior to the
crashes of 1929, 1962 and 1987. The existence of a ``slow'' crash in 1962 was
before these results unknown to us and the identification of this crash
naturally strengthens our case. (A systematic unpublished prediction scheme 
used on 20 years of the Hang-Seng index also detected a crash unknown to
us in the second half of 1981, see section \ref{alarm}). Thus, the
results from fitting the surrogate data sets generated from the real
stock market index show that fits, which in terms of the fitting parameters
corresponds to the three crashes of 1929, 1962 and 1987, are not 
likely to occur
``accidentally''. Actually,
Feigenbaum also notes ``Both Feigenbaum and Freund and Sornette et 
al. also looked
at randomly selected time widows in the real data and generally found 
no evidence
of log-periodicity in these windows unless they were looking at a time
period in which a crash was imminent''. Is this not an important 
out-of-sample test?
But Feigenbaum then turns a deaf ear to these results, adding ``[...] it is
still not clear what, if any, conclusions can be drawn from them''!

Feigenbaum [2001] seriously misquotes our numerical experiment
[Johansen et al., 2000b] testing whether
the null hypothesis that GARCH(1,1) with Student-distributed noise
could explain the presence of log-periodicity. The statement that
we ``found no evidence in the simulated data of a linkage between fits
satisfying our criteria and ensuing crashes, in contrast to the linkage
we report in the real data'' misrepresents our goals and results.
In the 1000 surrogate data sets of length 400-weeks generated using the
GARCH(1,1) model with Student-distributed noise and analyzed as for 
the real crashes,
we found only two 400-weeks windows which qualified.
This result corresponds to a confidence level of $99.8\%$ for rejecting
the hypothesis that GARCH(1,1) with Student-distributed noise can generate
meaningful log-periodicity. There is no reference to a crash, the question is
solely to test if log-periodicity of the strength observed before 1929 and 1987
can be generated by one of the standard benchmark of financial time series used
intensively both by academics and practitioners. If in addition, we add
that the two spells of significant log-periodicity generated in the simulations
using GARCH(1,1) with Student-distributed noise were not following by crashes,
then the case is even stronger for concluding that real markets exhibit
behaviors that are
dramatically different from the one predicted by one of the most 
fundamental benchmark
of the industry! We note also that Feigenbaum's remark that ``the 
criteria used by Johansen
et al. would have rejected our best fit for the precursor to the 1987 crash''
is misplaced: Feigenbaum and Freund [1996] used the simpler 
one-frequency formula
on a much narrower time window and the two procedures can thus not be compared.

Feigenbaum's remark, that ``all these simulation results are weakened 
by the fact
that each experiment rules out only one possible data generating process'',
is a truism: no truth is ever demonstrated in science; the only
thing that can be done is to construct models and reject them at a given
level of statistical significance. Those models, which are not rejected
when pitting them
against more and more data  progressively acquire the
status of theory (think for instance of quantum mechanics which is again and
again put to tests). In the present context, it is clear that, in a 
purist sense,
we shall never be able to ``prove'' in an absolute sense the existence of a 
log-periodicity genuinely
associated with specific market mechanisms. The next best thing we can do is
to take one by one the best benchmarks of the industry and test them to
see if they can generate the same structures as we document. If more and more
models are unable to ``explain'' the observed log-periodicity, this means that
log-periodicity is an important ``stylized'' fact that needs to be understood.
This is
what Feigenbaum and Freund [1996] have done using the random walk 
paradigm and this is
what we have done using the GARCH(1,1) with Student-distributed noise 
industry standard.
In his section 4, Feigenbaum changes his mind: ``our main concern is whether
any apparent log-periodicity can be accounted for by a simple specification of
the index's behavior. A random walk with drift would satisfy this purpose,
and that is what we use.'' Here, we can agree heartily!
It would of course be interesting to test more sophisticated models 
in the same way.
However, we caution that rejecting one model after another will never 
prove definitively that
log-periodicity exists. This is outside the realm of statistical and 
econometric analysis.

Feigenbaum's statement that
``the frequency of crashes in Johansen et al.'s Monte Carlo simulations was
much smaller than the frequency of crashes in real data, so their 
Data Generating Process
obviously does not adequately capture the behavior of stock prices 
during a crash''
falls in the same category. His sentence implies that our test is 
meaningless while
this is actually the opposite: again, if the most-used benchmark of 
the industry
is incapable of reproducing the observed frequency of crashes, this 
indeed means
that there is something to explain that may require new concepts and 
methods, as
we propose.

In his section 4, Feigenbaum examines the
first differences for the logarithm of the S\&P 500 from
1980 to 1987 and finds that he cannot reject
the log-periodic component $\Delta f_2$ at the $95\%$ level:
in plain words, this means that the probability that the log-periodic
component results from chance is about or less than $0.05$. This is an
interesting result that complements our previous investigations.
In contrast,
he finds that the pure power tern $\Delta f_1$ has a probability of 
about $30\%$
to be generated by chance alone and he attributes this to the weakness
of its justification. We disagree: in our derivation [Sornette and 
Johansen, 1997],
the $f_1$ term plays the same role as the $f_2$ term in the 
Landau-Ginzburg expansion.
The origin of the lack of strong statistical significance of this term
$\Delta f_1$ results rather, in our opinion, from the intrinsic difficulty
in quantifying a trend and an acceleration in very noisy data. This 
was for instance shown in another noisy data associated with the Kobe 
earthquake [Johansen et al., 1996] for which a pure power law could absolutely 
not be fitted without fixing $t_c$ (the fitting algorithm could not converge 
to a $t_c > t_{last}$ where $t_{last}$ is the timing of the last data point) 
while an exponential or a log-periodic power law could be fitted to the data.

Feigenbaum concludes his statistical section 4 by suggesting that an integrated
process, like a random walk which sums up random innovations over time, can
generate log-periodic patterns. Actually, Huang et al. [2000a] tested
specifically the following problem: under what circumstances can an integrated
process produce spurious log-periodicity? The answer obtained after lengthy and
thorough Monte-Carlo tests is twofold. (a) For approximately regularly sampled
time series as is the case of the financial time series,
taking the cumulative function of a noisy log-periodic function {\it destroys}
the log-periodic signal! (b) Only when sampling rates increase exponentially or
as a power law of $t_c-t$ can spurious log-periodicity in integrated 
processes be
observed.

\section{The issue of prediction}

\subsection{Addressing Feigenbaum's criticisms}

The determination of the parameters in a power law $B (t_c-t)^{\beta}$ is 
very sensitive
to noise and to the distance from $t_c$ of the data used in the estimation.
This is well-known by experimentalists and numerical scientists
working on critical phenomena in condensed matter physics
who have invested considerable efforts 
in developing
reliable experiments that could probe the system as close as possible to the
critical point $t_c$, in order to get reliable estimations of $t_c$ 
and $\beta$.
A typical rule of thumb is that an error of less than
$1\%$ in the determination of $t_c$ can lead to tenth of percent errors in
the estimation of the critical exponent $\beta$. We stress here that the
addition of the log-periodic component improves significantly the 
determination of
$t_c$ as the fit can ``lock-in'' on the oscillations. This is what we have
repeatedly shown in our various applications of log-periodicity (see 
first section).

It is in this context that the following report of Feigenbaum should 
be analyzed:
``excluding the last year of data, the log-periodic component is
no longer statistically significant. Furthermore, the best fit in this
truncated data set predicts a crash in June of 1986, shortly after the data
set ends but well before the actual crash''. This is not surprising in view
of what we just recalled: it is as if a worker on critical phenomena
was trying to get a reliable estimation of $t_c$ and $\beta$ by thrashing
the last $15\%$ of the data, which are of course the most relevant. 
But trying to
test log-periodicity, one cannot but conclude that Feigenbaum is throwing
the baby out with the bath. To be constructive, we offer the following clues.
(i) We would have rejected the fit shown in table 3 [Feigenbaum, 2001] as
the value of $\Delta t=1285$ years is meaningless. We would have concluded that
a prediction so much in advance is not warranted. (ii) In those cases where
we either made an ex-ante announcement [Johansen and Sornette, 1997; 1999c]
or reported afterward a prediction
experiment [Johansen and Sornette, 2000b; 2000c] (see also 
[Vandewalle et al., 1998b]),
the predictions used
data ending within a few months at most from the actual $t_c$. Again, 
it is all too easy
to reject an hypothesis by naively applying it outside its domain of
application. It is true that we had not stated
a precise methodology for predicting crashes or strong corrections on the
basis of log-periodic signals in our previous papers.
This should not be understood as meaning
that the model should be applied blindly and that it should be rejected
as soon as there is a failure! Our purpose has been and is still
to explore the possibility of a novel remarkable signatures of the collective
behavior of investors. Our work suggests the existence of a cooperative
behavior of investors, leading to a fundamental ripening of the 
markets toward an
instability. Our work suggests
the possibility of using the proposed framework to predict when the market will
exhibit a crash/major correction. Our analysis not only points to a
predictive potential but also that false alarms are difficult to avoid
due to the underlying nature of speculative bubbles. This should not however
be the main emphasis as this will distract from the fundamental issues.

\subsection{The example of the Hong Kong crashes} \label{alarm}

\begin{figure}
\begin{center}
\parbox[l]{8.5cm}{
\epsfig{file=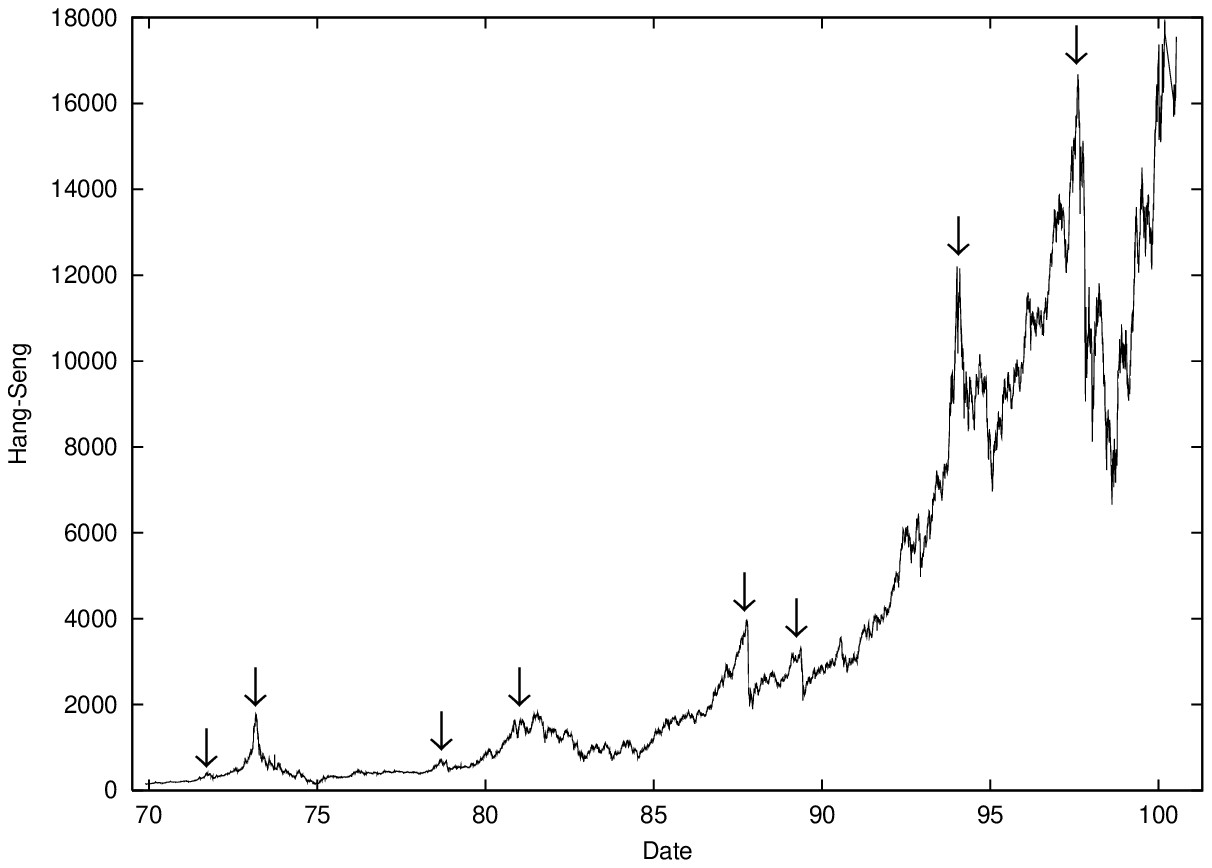,height=8cm,width=8.5cm}}
\hspace{5mm}
\parbox[r]{8.5cm}{
\epsfig{file=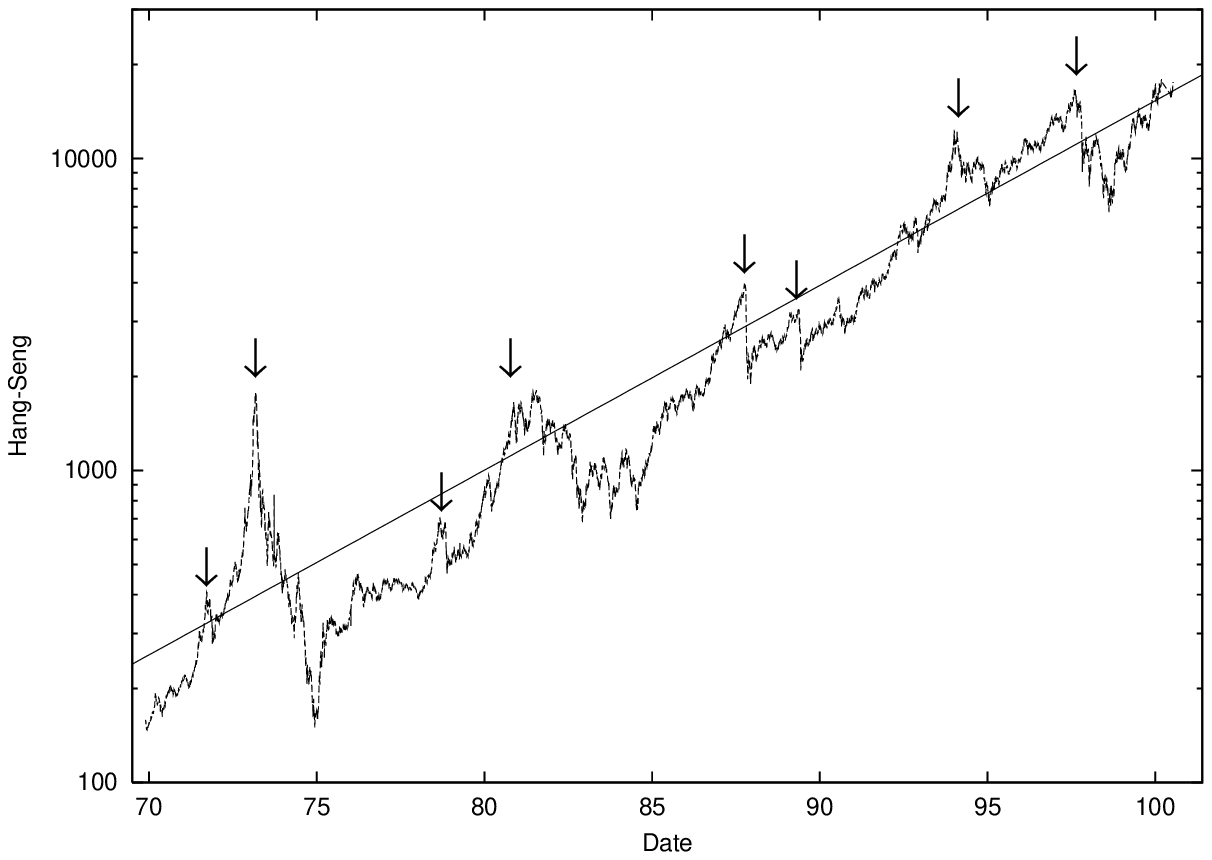,height=8cm,width=8.5cm}}
\caption{\protect\label{scanhk} The Hang-Seng composite index of the Hong Kong 
stock market from Nov. 1969 to Sept. 1999. First panel: linear scale; 
second panel: logarithmic scale in the vertical axis. The culmination
of the bubbles followed by strong corrections of crashes are indicated by the symbols
`fl' and correspond to the times Oct. 
1971, Feb, 1973, Sept. 1978, Oct. 1980, Oct. 1987, April 1989, Jan. 1994 and Oct. 1997. 
The second panel with the index in 
logarithmic scale shows that it has grown exponentially on average at the rate
of $\approx 13.6\%$ per year represented by the straight line corresponding to 
the best exponential fit to the data. Eight large bubbles (among them 5 are
very large) can be observed 
as upward accelerating deviations from the average exponential growth. 
There are also smaller structures that we do not consider here.
}
\end{center}
\end{figure}

In order to test the potential of the log-periodic
power law formula and our numerical procedures with 
respect to predicting crashes and large corrections, the Hang-Seng index was fitted 
from $\approx 1980$ to $\approx 1999$ in segments of 
$\approx 1.5$ years, see figure \ref{scanhk}. For each new fit, the 
start and end points of the 
time interval used in the fit were simultaneously moved forward in time in 
steps of 5 points corresponding to a fit every new trading week. 

Previous results [Johansen et al., 1999] have shown that, for the major 
financial markets, the values of the two physical variables, the exponent 
$\beta$ and the angular log-frequency $\omega$, are subject  to constraints. 
Specifically, we have found that the {\em price} prior to large 
crashes and corrections obeys the log-periodic power law
formula with $0.1 < \beta < 0.8$ and $6 < \omega <  9$ (with the possible
existence of harmonics). The constraint on $\omega$ is documented in 
figure \ref{lomballhk}, which shows the Lomb periodogram of the log-periodic
component of the Hang-Seng price index for the eight bubbles observed in 
figure \ref{scanhk}, ending in Oct. 1971, in Feb, 1973, in Sept. 1978, in Oct. 
1980, in April 1989, in Nov. 1980, in Oct. 1987, in Jan. 1994 and in Oct. 1997.
It is striking that the log-periodic spectra of all eight bubbles peak 
basically at the same log-frequency $f = 1.1 \pm 0.3$ ($\pm 0.2$ if we exclude
the first two bubbles), corresponding to an angular frequency $5.0 \leq 
\omega=2 \pi f \leq 8.8$. In addition, sub-harmonics (half log-frequency) and 
harmonics $2f$ can be seen. For completeness and for the sake of comparison, 
figure \ref{lomballws} shows the log-periodic spectra for all the major bubbles
ending in crashes on the Dow Jones and SP500 indices in the twentieth century. 
Observe that the sub-harmonics (half log-frequency) and two harmonics $2f$ and 
$3f$ are quite strong in a few of the data sets. Figure  \ref{lomballfx} shows 
the log-periodic spectra for the major bubbles on currencies, with similar 
conclusions.

For prediction purpose, an additional technical constraint we may impose on a 
fit in order for it to qualify as an alarm is that $B$ should be negative 
(because we are looking for a bubble ending in a crash. Specifically, this 
criterion disqualifies Feigenbaums ``1974 case''.). Last, a fit should not
qualify as an alarm if it does not belong to a ``cluster'' of qualifying fits, 
{\it i.e.}, a relative small change in the start and end dates for the 
time-interval used should not alter the result.

It is necessary to decide on a horizon for the search with respect to the
timing of the end of the bubble. Since we are looking for an impending crash
along the time series, 
the search for $t_c$ started between one week and $\approx 0.2$ years after 
the last date in the data set. However, $t_c$ is not fixed and is
one of the parameters determined in the fitting procedure.

Alarms was produced in the following nine time intervals containing the date of
the last point used in the fit:

\begin{enumerate}

\item 1981.60 to 1981.68. This was followed by a $\approx 30 \%$ decline.

\item 1984.36 to 1984.41. This was followed by a $\approx 30 \%$ decline.

\item 1985.20 to 1985.30. False alarm.

\item 1987.66 to 1987.82. This was followed by a $\approx 50$ \% decline.

\item 1989.32 to 1989.38. This was followed by a $\approx 35$ \% decline.

\item 1991.54 to 1991.69. This was followed by a $\approx 7 \% 1$ single day decline.
Considered a false alarm, nevertheless.

\item 1992.37 to 1992.58. this was followed by a $\approx 15 \%$ decline. 
This is a marginal case.

\item 1993.79 to 1993.90. This was followed by a $\approx 20 \%$ decline.
This can also be considered as a marginal case, if we want to be
conservative.

\item 1997.58 to 1997.74. This was followed by $\approx 35 \%$ decline.

\end{enumerate}

We end up with two to four false alarms and seven to five correct predictions. 
Note that the alarms 
do not depend on whether $t_c$ is correctly estimated or not. In order to 
include the timing of the crash/large correction, the size of the data 
interval must be adapted in each particular case. 
In addition, there is a clearly identifiable
miss by our procedure, namely the $\approx 25$ \% decline of July-August 1990.
The large majority of the other large declines not identified by the procedure
belongs to the turbulent periods following the ones identified.

\subsection{Statistical confidence of the crash ``roulette''  \label{crashroul}}

What is the statistical significance of a prediction scheme that
made at least five correct predictions, issued at most four false alarms and 
missed at least one event?

To formulate the problem 
precisely, we divide time in bimonthly intervals and 
ask what is the probability that a crash or strong
correction occurs in a given two-month interval.
Let us consider $N$ such two-month intervals. The 
period over which we carried out our analysis
goes from Jan. 1980 to Dec. 1999, corresponding to $N=120$ two-months.
In these $N=120$ time intervals, $n_c=6+x$ crashes occurred while $N-n_c=114-x$
bimonthly period were without crash, where we allow for $x > 0$ to check
the sensitivity of the statistical confidence to the definition and
thus detection of crash or strong corrections.
 Over this 20-year time interval, we made
$r=9$ predictions and $k=5$ of them where successful while $r-k=4$ 
were false alarms. What is the probability $P_k$ to have such a success
from chance?

This question has a clear mathematical answer and reduces to a well-known
combinatorial problem leading to the so-called Hypergeometric 
distribution. The solution of this problem is given in Appendix B for
completeness.

In the case of interest here, we plug in the formula given in the Appendix
B the following numbers: the number
of bimonthly periods is $N=120$, the number of real crashes is $n_c=6+x$,
the number of correct predictions is (conservatively) $k=5$,
$N-n_c=114-x$, the total number of issued prediction is
$r=9$ and the number of false alarms is $r-k=4$. For $x=0$, the corresponding
probability for this scenario to result from chance is
$P_{k=5}(x=0)=3.8 \cdot 10^{-6}$. 

If we add the possibility for another missed event, this leads to
a number of real crashes equal to $n_c=7$,
the number of correct predictions is still (conservatively) $k=5$, 
the number of bimonthly periods without crash is
$N-n_c=113$, the total number of issued prediction is still
$r=9$ and the number of false alarms is still $r-k=4$. The corresponding
probability for this scenario to result from chance is 
$P_{k=5}(x=1)=1.3 \cdot 10^{-5}$. 

Even with $x=5$ additional missed events, the probability that
our five successful predictions result from chance remains very low.
In this case, $n_c=11$,
the number of correct predictions is still (conservatively) $k=5$, 
the number of bimonthly periods without crash is
$N-n_c=109$, the total number of issued prediction is still
$r=9$ and the number of false alarms is still $r-k=4$. The corresponding
probability for this scenario to result from chance is 
$P_{k=5}(x=5)=2.5\cdot 10^{-4}$. 

We conclude that this track record, while
containing only a few cases, is highly suggestive of a real
significance.  
We should stress that this contrasts with the naive view
that five successes and four false alarms and a few missed events, would
corresponds approximately to one chance in two to be right, giving the 
impression that the prediction skill is no better than deciding
that a crash will occur by random coin tosses.
This conclusion would be very naive because it forgets an essential element of
the forecasting approach, which is to identify a (short) time window
(two month) in which
a crash is probable: the main difficulty in making a prediction is indeed
to identify the few bimonthly periods among the 
$120$ in which there is the risk of a crash. We note that these
estimations remain robust when changing the time windows by large
variations.

\subsection{Using crash predictions}

In evaluating predictions and their impact on (investment) decisions,
one must weight the relative cost of false alarms with respect to the
gain resulting from correct predictions. The Neyman-Pearson diagram, also 
called the decision quality diagram, is used in optimizing decision
strategies with a single test statistic. The assumption is that samples of 
events or probability density functions are available both for the correct
signal (the crashes) and the background noise (false alarms); 
a suitable test statistics is then sought which  
distinguishes between the two in an optimal fashion. Using a given test statistics (or
discriminant function), one can introduce a cut which separates an acceptance
region (dominated by correct predictions) from a rejection region (dominated by
false alarms). The Neyman-Pearson diagram plots contamination (mis-classified
events, {\it i.e.}, classified as predictions which are thus false alarms) 
against losses (mis-classified
signal events, {\it i.e.}, classified as background or failure-to-predict), 
both as fractions of the total sample.
An ideal test statistic corresponds to a diagram where the ``Acceptance
of prediction'' is plotted as a function of the ``acceptance of false alarm''
in which the acceptance is close to $1$ for the real signals, and close
to $0$ for the false alarms. Different strategies are possible:
a ``liberal'' strategy favors minimal loss (i.e. high
acceptance of signal, {\it i.e.}, almost no failure to catch the real events
but many false alarms), a ``conservative'' one favors minimal contamination 
({\it i.e.}, high purity of signal and almost no false alarms).
Molchan [1990, 1997] has reformulated this Neyman-Pearson diagram into an 
``error diagram''
which plots the rate of failure-to-predict (the number of missed
events divided by the total number of events in the total time interval)
as a function of the rate
of time alarms (the total time of alarms divided by the total time, in other
words the fraction of time we declare that a crash is looming).
The best predictor corresponds to a point close to the origin in this
diagram, with almost no failure-to-predict and with a small fraction of time
declared as dangerous: in other words, this ideal strategy misses no event
and does not declare false alarms! The correspondence with the
Neyman-Pearson diagram is 
acceptance of signal $=$ 1/(1-rate of failure) and
acceptance of background $=$ 1/(1-rate of false alarms). 
These considerations teach us that making a prediction is one thing, using
it is another which corresponds to solving a control optimization problem
[Molchan, 1990; 1997].

\begin{figure}
\begin{center}
\epsfig{file=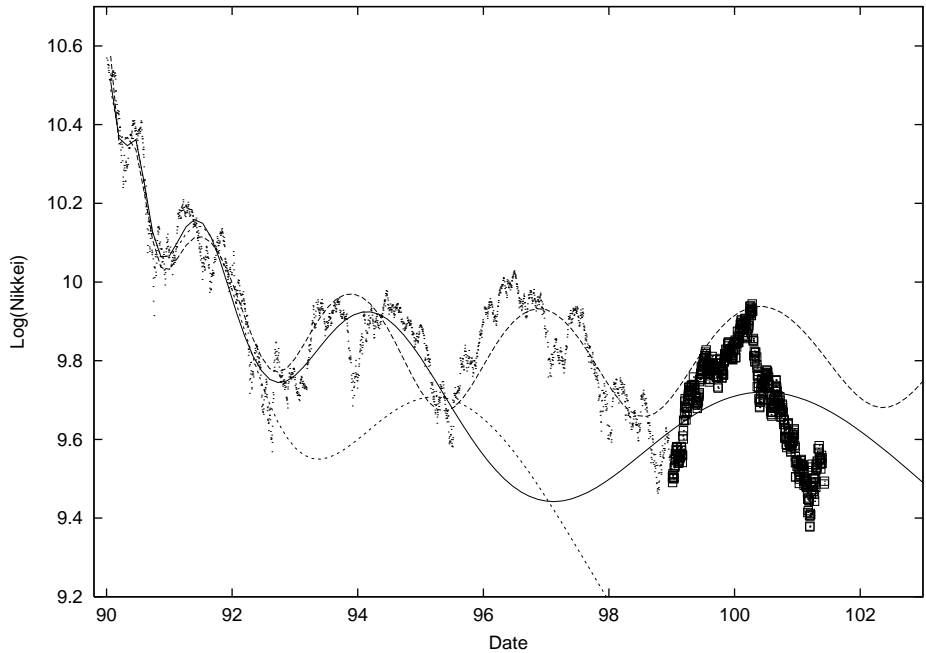}
\caption{\protect\label{Nikkeipredic} 
In early Jan. 1999, Johansen and Sornette (1999c) fitted the Japanese Nikkei index
from 1 Jan 1990 to 31 Dec. 1998 with three log-periodic formula
corresponding to successive improvements in a Landau-expansion of the
renormalization group formulation of a bubble: linear log-periodic
formula (dotted line), second-order nonlinear log-periodic formula (continuous line)
and third-order nonlinear log-periodic formula (dashed line). This last
curve provided a good fit to the data from 1 Jan 1990 to 31 Dec. 1998. We 
therefore used it to extrapole to the future. Our prediction announced in Jan. 1999
predicted that
the Japanese stock market should increase by about $50\%$ 
as the year 2000 was approached, would culminate and then decrease again.
In this figure, the value of the Nikkei is represented as the dots for the data
used in the fits and as squares for the out-of-sample data (Jan. 1999 to May 2001).
This figure is as in (Johansen and Sornette, 1999c) except for the squares starting
3rd Jan. 1999 which
represents the realized Nikkei prices since the prediction was issued. The prediction
turned out to correctly predict the level and time of the maximum 
(20800 in April 2000). Since then, the quality of the extrapolation has deteriorated
even if the existence of a downward trend leading to another minimum was expected.
The observed minimum 
below 13000 in early March 2001 anticipated the extrapolation of the dashed curve.
More nonlinear terms should have been included to extend the prediction further
in the future, as can be seen from the increasing improvements of the log-periodic
formulas as their nonlinear order increases. See [Johansen and Sornette, 2000c] for
a quantitative evalution of this prediction.
}
\end{center}
\end{figure}

Indeed, suppose that a crash prediction is issued stating that a
crash will occur $x$ weeks from now. At least three different scenarios
are possible [Johansen and Sornette, 2000b]:
\begin{itemize}

\item Nobody believes the prediction which was then futile and, assuming
that the prediction was correct, the market crashes.
One may consider
this as a victory for the ``predictors'' but as we have experienced in relation
to our quantitative prediction of the change in regime of the Nikkei index 
[Johansen and Sornette, 1999c] (see figure \ref{Nikkeipredic} for an up-to-date
assessment of this prediction), this would only be considered by some critics 
just another ``lucky one'' without any statistical significance (see
[Johansen and Sornette, 2000c] for an alternative Bayesian approach).

\item Everybody believes in the warning, which causes panic and the market
crashes as consequence. The prediction hence seems self-fulfilling.

\item Enough believe that the prediction {\it may} be correct and take
preemptive actions which make he steam
go off the bubble. The prediction hence disproves itself.

\end{itemize}

None of these scenarios are attractive from a practical point of view.
In the first two, the crash is not
avoided and in the last scenario the prediction disproves itself and as
a consequence the theory looks unreliable. This seems to be the inescapable
lot of scientific investigations of systems with learning and reflective
abilities, in contrast with the usual inanimate and unchanging physical laws
of nature. Furthermore, this touches the key-problem of scientific
responsibility. Naturally, scientists have a responsibility to publish their
findings. However, when it comes to the practical implementation of those
findings in society, the question becomes considerably more complex,
as numerous historical and modern instances have shown us.

\section{Are large drawdowns outliers?}

\subsection{A first cautionary remark on testing drawdown outliers}

A drawdown is defined as a
persistent decrease in the price over consecutive days. A drawdown
is thus the cumulative loss from the
last maximum to the next minimum of the price. Drawdowns embody
a rather subtle dependence
since they are constructed from runs of the same sign variations. Their
distribution thus captures the way successive drops can influence 
each other and
construct in this way a persistent process.
This persistence is not measured by the
distribution of returns because, by its very definition, it forgets about
the relative positions of the returns as they unravel themselves as a function
of time by only counting their frequency. This is not detected either by
the two-point correlation function, which measures an {\it average}
linear dependence over the whole time series, while the dependence may only
appear at special times, for instance for
  very large runs, a feature that
will be washed out by the global averaging procedure.

\begin{figure}
\begin{center}
\epsfig{file=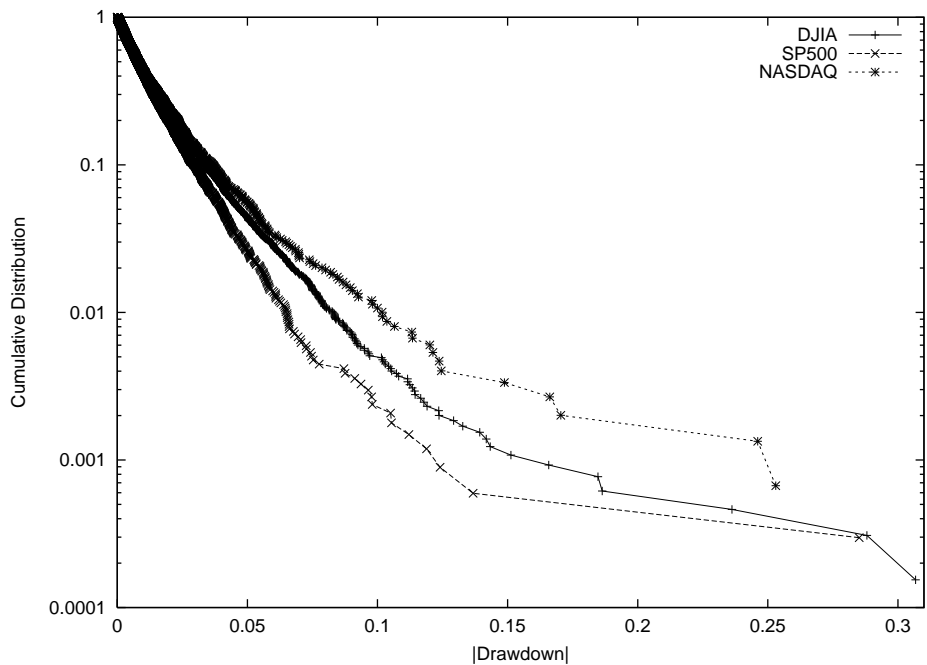}
\caption{\protect\label{cumudistrib} Complementary cumulative distribution of 
the (absolute value of the) drawdowns for the Dow Jones Industrial Average index
($+$) from 1900.00 to 2000.34, S\&P500 index ($\times$) from 1940.91 to 2000.34 and 
the NASDAQ index ($\ast$) from 1971.10 to 2000.30. The ordinate is in 
logarithmic scale while the abscissa shows the absolute value of the drawdowns.
The total number of drawdowns for the three index are $6486$, $3363$ and 
$1495$, respectively.
}
\end{center}
\end{figure}

Figure \ref{cumudistrib} shows the complementary cumulative distribution of 
the (absolute value of the) drawdowns for the Dow Jones Industrial Average index
from 1900.00 to 2000.34, of the S\&P500 index from 1940.91 to 2000.34 and 
of the NASDAQ index from 1971.10 to 2000.30.
The question is whether the upward curvature seen for these three indices
qualifies a change of regime.

In his section 5, Feigenbaum attempts to show that our detection
[Johansen and Sornette, 1988b; 2001b]
of two populations of drawdowns does not survive his testing
procedure. His argument is twofold: (i) the exponential and
stretched-exponential null hypotheses are rejected at the $95\%$ 
confidence level
and (ii) a modified stretched exponential distribution appears to fit
the distribution of drawdowns, including the 1987 crash. This second point
is misleading as his table 10 shows only that the coefficients of his model
are necessary, not that it is a good model. Following his own
procedure, he should have included an additional
term such as a $d^{3/2}$ in the exponential and asked whether the 
null-hypothesis
that the coefficient of this new term is zero is rejected. Furthermore, the
t-test is too weak in general to detect the impact of one or a few 
outliers. Thus, the fact that
his inclusion of the 1987 crash drawdown does not change significantly
the statistics does not mean anything.

Actually, testing for ``outliers'' or more generally for a change of
population in a distribution is a subtle problem, that
escaped the attention of even some of our cleverest colleagues for some time
and is still overlooked by many others. This subtle point is that
the evidence for outliers and extreme events does not require
and is not even synonymous in general with the existence of
a break in the distribution of the drawdowns. Let us illustrate this
pictorially by borrowing from another domain of active scientific
investigation, namely the search for the understanding of the complexity
of eddies and vortices in turbulent hydrodynamic flows, such as
in mountain rivers or in the weather. Since
solving the exact equations of these flows does not provide
much insight as the results are forbidding,
a useful line of attack has been to simplify the problem
by studying simple toy models, such as so-called
``shell'' models of turbulence, that are believed to capture the
essential ingredient of these flows, while being amenable
to analysis. Such ``shell'' models replace the three-dimensional
spatial domain by a series of uniform onion-like spherical layers
with radii increasing as a geometrical series $1, 2, 4, 8, ..., 2^n$
and communicating with each other mostly with nearest neighbors.

As for financial returns, a quantity of great interest
is the distribution of velocity variations between two instants at
the same position or between two points simultaneously. Such a
distribution for the square of the velocity variations has been
calculated  [L'vov et al., 2001] and exhibits an approximate
exponential drop-off as well as a co-existence with larger fluctuations, 
quite reminiscent of our findings in finance 
[Johansen and Sornette, 1988b; 2001b] and of figure 11 in [Feigenbaum, 2001].
Usually, such large fluctuations are not considered to be
statistically significant and do not provide any specific insight.
Here, it turns out that it can be shown that these large
fluctuations of the fluid velocity correspond to intensive
peaks propagating coherently over several shell layers with a
characteristic bell-like shape, approximately independent of
their amplitude and duration (up to a re-scaling of their size
and duration). When extending these observations to very long
times so that the anomalous fluctuations can be sampled much better, one gets
a continuous distribution [L'vov et al., 2001]. Naively, one would 
expect that the
same physics apply in each shell layer (each scale) and, as a consequence,
the distributions in each shell should be the same, up to a change of 
unit reflecting
the different scale embodied by each layer. It turns out that the three curves
for three different shells
can indeed by nicely collapsed, but only for the small velocity 
fluctuations, while
the large fluctuations are described by very different heavy tails.
Alternatively, when one tries to collapse the curves in the region of the large
velocity fluctuations, then the portions of the curves close to the origin
are not collapsed at all and are very different.
The remarkable conclusion is that the distributions
of velocity increment seem to be composed of two regions, a region of so-called
``normal scaling'' and a domain of extreme events.
The theoretical analysis of L'vov et al. [2001] further substantiate
the fact that the largest fluctuations result from a different mechanism.

Here is the message that comes out of this discussion: the concept of 
outliers and of
extreme events does not rest on the
requirement that the distribution should not be smooth. Noise and
the very process of constructing the distribution
will almost always smooth out the curves. What is found by L'vov et al. [2001]
is that the distribution is made of two different populations, the body and
the tail, which have different physics, different
scaling and different properties. This is a clear demonstration that this model
of turbulence exhibits outliers in the sense that there is a well-defined
population of very large and quite rare events that punctuate the dynamics and
which cannot be seen as scale-up versions of the small fluctuations.

As a consequence, the fact that the distribution of small events might show
up some curvature or continuous behavior does not tell anything against
the outlier hypothesis. It is essential to keep this point in mind in
looking at the evidence presented below for the drawdowns.

Other groups have recently presented supporting evidence that 
crash and rally days significantly differ in their statistical properties from the 
typical market days  (Lillo and Mantegna, 2000).
For instance, Lillo and Mantegna investigated the return distributions of an ensemble 
of stocks simultaneously traded in the New York Stock Exchange (NYSE) during
market days of extreme crash or rally 
in the period from January 1987 to December 1998. Out of
two hundred distributions of returns, one for each 
of two hundred trading days where the ensemble of returns is constructed over the whole
set of stocks traded on the NYSE, anomalous large widths
and fat tails are observed specifically
on the day of the crash of Oct. 19 1987, as well as during a few other
turbulent days. Lillo and Mantegna document another 
remarkable behavior associated with crashes
and rallies, namely that the distortion of the distributions of returns
are not only strong in the tails describing large moves but also in their center.
Specifically, they show that the overall shape of the distributions
is modified in crash and rally days. 
Closer to our claim that markets develop precursory signatures of bubbles
of long time scales, Mansilla (2001) has also shown, using a measure of
relative complexity, that time sequences corresponding to ``critical''
periods before large market corrections or crashes
have more novel informations with respect to the whole price time series than those
sequences corresponding to periods where nothing happened. The conclusion is that the
intervals where no financial turbulence is observed, that is, where the markets works fine,
the informational contents of the (binary-coded) price time series is small. 
In contrast, there seems to be significant information in the price time series
associated with bubbles. This finding is consistent with the appearance of 
a collective herding behavior modifying the texture of the price time series compared
to normal times.

In order to make further progress, we present three statistical tests that
complement each other and strengthen our claim that very large
drawdowns are outliers.

\subsection{Surrogate data analysis}

Reshuffling the distributions of returns provides a powerful tool for
qualifying the existence of higher order correlations in the drawdown
distributions. We have reshuffled 10,000 times the daily 
returns of the Nasdaq Composite
index since its establishment 1971 until 18 April 2000. We have thus
generated 10,000 synthetic data sets [Johansen and Sornette, 2000b]. 
This procedure means 
that the synthetic data will have {\it exactly} the same distribution of 
daily returns as the real data. However, higher order correlations apparently
present in the largest drawdowns are destroyed by the reshuffling. This
surrogate data analysis of the distribution of drawdowns has the advantage
of being {\it non-parametric}, {\it i.e.}, independent of the quality of fits
with a model such as the stretched exponential or the power law. The draw-back
is that this kind of bootstrap analysis is ``negative'', {\it i.e.}, we use
these tests to {\it reject} a given null-hypothesis, not to {\it confirm} a
given hypothesis.

Out of the 10,000 synthetic data sets, 776 had a single drawdown larger
than $16.5\%$, 13 had two drawdowns larger than $16.5\%$, 1 had three drawdowns 
larger than $16.5\%$ and none had 4 (or more) drawdowns larger than
$16.5\%$ as in the real data. This means that given the distribution of
returns, by chance we have a $\approx 8\%$ probability of observing a drawdowns 
larger than $16.5\%$, a $\approx 0.1\%$ probability of observing two
drawdowns larger than $16.5\%$ and for all practical purposes zero
probability of observing three or more drawdowns larger than $16.5\%$.
Hence, the probability that the largest four drawdowns observed for the
Nasdaq could result from chance is less than $0.01\%$.
As a consequence we are lead to conclude that the largest market events are
to be characterized by the presence of higher order correlations in contrast 
to what is observed during ``normal'' times.

\subsection{GARCH Analysis}

Another approach is to use the GARCH(1,1) model discussed above
with Student distribution of the noise with $4$ degrees of freedom
fitted to the Dow Jones Industrial Average. The model allows us to
generate synthetic time series that capture the main
stylized facts of financial time series. The appendix recalls
for completeness how these synthetic time series are generated.

\begin{figure}
\begin{center}
\epsfig{file=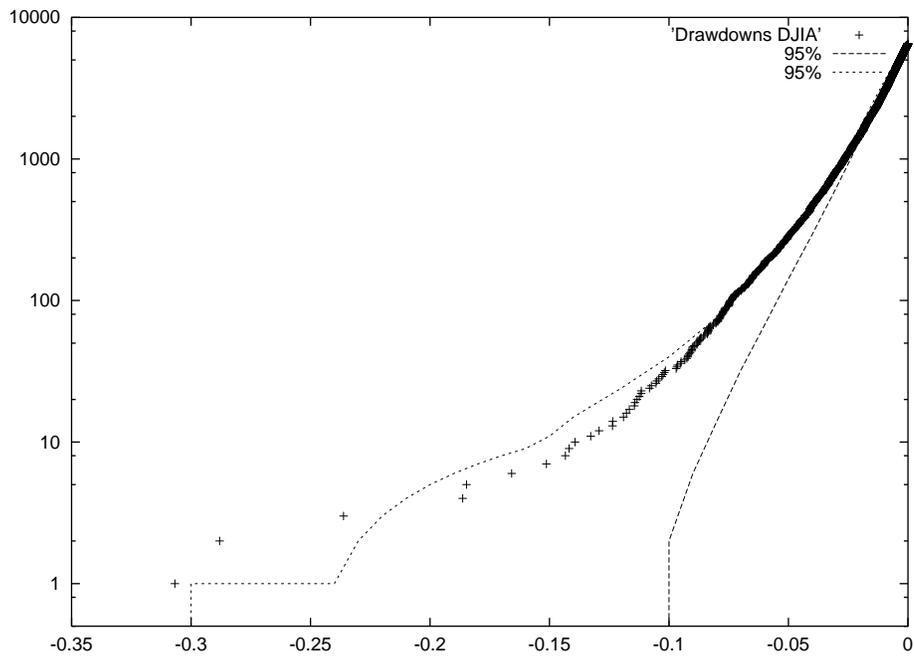}
\caption{\protect\label{trailgarchconfcr}
The two dashed lines are
defined such that 99\% of the drawdowns of synthetic GARCH(1,1) with 
noise student distribution with $4$ degrees of 
freedom are within the two lines. The symbols $+$ 
represent the cumulative 
distribution of drawdowns for the Dow Jones index. The ordinate
is in logarithmic scale while the abscissa shows the drawdowns:
for instance, $-0.30$ corresponds to a drawdown of $-30\%$.
}
\end{center}
\end{figure}

>From such synthetic
price time series, we construct the distribution of drawdowns
following exactly the same procedure as in the analysis of the real 
time series. Figure \ref{trailgarchconfcr} shows two continuous lines
defined such that 95\% of the drawdowns of synthetic GARCH(1,1) with 
noise student distribution are within the two lines: there is thus a 2.5\%
probability that a drawdown in a GARCH(1,1) time series 
with Student distribution of the noise with $4$ degrees of freedom falls
above the upper line or below the lower line. Notice that the 
distribution of drawdowns from the synthetic GARCH model is approximately
exponential or slightly sub-exponential for drawdowns up to about $10\%$ 
and fits well the empirical drawdown distribution
shown as the symbols $+$ for the Dow Jones index. However,
the three largest drawdowns are clearly above the upper line. We
conclude that GARCH(1,1) dependencies, notwithstanding
the correct fat-tailness of the distribution of returns, 
can not (fully) account for the 
dependence observed in real data, in particular in the special
dependence associated with very large drawdowns. This illustrates
that one of the most used benchmarks in finance fails to match the data
with respect to the largest drawdowns.

This novel piece of evidence, adding upon the previous rejection of the null
hypothesis that reshuffled time series exhibit the same drawdowns as
the real time series, strengthens the claim that large drawdowns
are outliers. We can of course never ``prove'' that large drawdowns
are outliers, but we can make the case stronger and stronger by rejecting
more and more null hypothesis. Up to now, we have rejected the reshuffled
data null hypotheses and the GARCH null hypothesis.

\subsection{Maximum Likelihood Analysis}

We now turn to another test formulated in the same spirit as done by 
Feigenbaum, which is aimed at the question whether there is a threshold 
quantile below which the null stretched exponential cannot be rejected and
beyond which it can. If this threshold quantile corresponds to say $5\%$ 
of the largest drawdowns, this suggests that most ($95\%$ to be specific) 
of the drawdowns are correctly modeled by the stretched exponential while 
the $5\%$ largest belong to a different distribution. We now present such a test.

First, let us recall the general framework of hypothesis testing within a 
parametric formulation. Suppose, the sample of drawdowns $X_1, ..., X_n$ 
has a probability density distribution (pdf) $p(x | a)$, where $a$ is some 
vector corresponding to the set of free parameters in the pdf. In our case,
this corresponds to taking the cumulative distribution
\be
P(x | a) = P(x=0) \exp \left[ - Bx^z + C x^{2z} \right]~,
\label{gjngnaq}
\ee
where $a = (B, C, z)$. The ``pure'' stretched exponential distribution 
corresponds hence to the case where $C=0$. The choice of this parameterization
(\ref{gjngnaq}) with a correction $C x^{2z}$ where the 
exponent is twice that of the first term in the exponential is taken
(1) to avoid introducing two additional parameters and (2) 
as the natural measure of a curvature in the log-linear plot
of $\ln P$ versus $x^z$ that would qualify the simple stretched 
exponential as a straight line.

In general, one considers two 
hypotheses corresponding to two sets of parameters $a = (a_1, ..., a_k)$:
\begin{enumerate}
\item $H_1$:  $a_1, ..., a_k$  belong to some $k$-dimensional interval $I_1$ 
(an infinite dimensional parameter space is possible as well). 

\item  $H_0$: one of the parameter $a_1, ..., a_k$ is equal to zero, say 
$a_1=0$ whereas the other parameters $a_2, ..., a_k$ can vary in the same 
$(k-1)D$-interval as in $H_1$ (more generally $H_0$  may assume that several
parameters are zero, say, $a_1=0, ..., a_m=0, m < k$). We denote this subset 
in parameter space as $I_0$. Evidently, the interval $I_1$ contains the 
interval $I_0$.
\end{enumerate}
 
Let us denote the maximum likelihood under $H_i$ as $L_i,  i = 0,1$:
\be 
L_0 = {\rm MAX}_0~\left[ p(X_1 | a)..p(X_n | a) \right]  \label{aaa}
\ee
where `MAX$_0$' is taken over $a$ in the parametric interval $I_0$, and
\be
L_1 = {\rm MAX}_1~\left[ p(X_1 | a)...p(X_n | a) \right]   \label{bbb}
\ee
where `''MAX$_1$'' is taken over $a$ in the parametric interval $I_1$.
By construction, $L_0 \leq L_1$ since adding one or several parameters
cannot decrease the quality of the fit to the data. A theorem by Wilks 
[Rao, 1965] states that, asymptotically as $n$ tends to infinity, the ratio
\be
T = -2 \log {L_0 \over L_1}    \label{testat}
\ee
is distributed as Chi-square with one degree of freedom (with $m$ degrees of 
freedom in the more general case of $m$ parameters with a fixed value).

In order to qualify the existence of outliers in the drawdown distributions
for the DJIA, SP500 and the NASDAQ indices, we have performed a maximum 
likelihood analysis tailored to estimate the significance of the curvature
seen in the distributions. We define
\bea 
P_{SE}(x) = A_{SE}(u) \exp \lp -Bx^z\rp   \label{2exps} \\
P_{MSE}(x) = A_{MSE}(u) \exp \lp -Bx^z + Cx^{2z}\rp   \label{2expsbis}
\eea
as two complementary cumulative distribution functions of drawdowns defined in
a given interval $\left[ 0, u\right]$. Hence the corresponding density distribution
functions are $p_{SE}(x) = -dP_{SE}(x)/dx$ and $p_{MSE}(x)=-dP_{MSE}(x)/dx$.
The subscripts SE and MSE stand for ``stretched exponential'' and
``modified stretched exponential'' respectively. Note that the normalizing factors 
$A_{SE}(u)$ and $A_{MSE}(u)$ are different and function of the upper cut-off $u$
since $p_{SE}(x)$ and $p_{MSE}(x)$ must be normalized to $1$ in the 
interval $\left[ 0, u\right]$. This normalization condition gives
$A_{SE}=1/[1- \exp \lp -Bu^z\rp]$ and $A_{MSE}=1/ [1- \exp \lp -Bu^z + Cu^{2z}\rp]$.

Technically, the maximum likelihood determination of the best 
parameters of the modified stretched exponential model $P_{MSE}(x)$ defined
by (\ref{2expsbis})
is done as a minimization of 
\bea
- \ln \lp L \rp = 
- \sum_{i=1}^{N} \ln p \lp X_i\rp = 
\sum_{i=1}^{N} \left( -\ln A_{MSE}(u) 
-\ln\lp Bz X_i^{z-1}-2zCX_i^{2z-1}\rp + BX_i^z - CX_i^{2z} \right)
\eea
with respect to the parameters $B, C$ and $z$,
using the downhill simplex minimization algorithm [Press {\it et al.~}, 1992],
and similarly for the stretched exponential model $P_{SE}(x)$ defined
by (\ref{2exps}). In order to secure that
the MLE indeed retains the parameter values of the global maximum, the 
downhill simplex minimization algorithm uses a wide range of start values
in the search.

Thus, for each distribution of drawdowns up to a given cut-off $u$, we perform 
a MLE of the parameters $B$ and $z$ for  $P_{SE}(x)$ defined
by (\ref{2exps}) and of $B, C$ and $z$ for $P_{MSE}(x)$ defined
by (\ref{2expsbis}). We ask whether we can reject the 
hypothesis that $C=0$. This test is based on the T-statistics comparing $H_0$ 
and $H_1$ defined by (\ref{testat}): if $T$ is large, $L_1$ is significantly 
larger than $L_0$ which means that adding the parameter $C \neq 0$ improves
significantly the quality of the fit. Hence, the pure stretched exponential 
model has to be rejected as mis-specified. If, on the other hand, $T$ is small, 
$L_1$ is not much larger than $L_0$ which means that adding the parameter 
$C \neq 0$ does not improve much the fit. The stretched exponential can be 
accepted as an adequate model, compared to the alternative specification 
allowing for a curvature in the $\ln N$ versus $x^z$ representation. This 
procedure will qualify the existence of outliers if the hypothesis $C=0$ 
cannot be rejected if $u$ is smaller than say $5\%$ while it can be rejected
for larger values of $u$.

Table \ref{tablenasdaq} shows the results of this test for the Nasdaq composite index.
We observe that the hypothesis $H_0$ that $C=0$ cannot be rejected for a cut-off
$u=3\%$ corresponding to $87\%$ of the total number of drawdowns. 
For $u=3\%$, the probability that $C= 0$ is
as high as $20.5\%$, which qualifies the stretched exponential model. However, for
$u=6\%$, i.e., $97\%$ of the total data set,
the SE model is rejected at the 95\% confidence level but not at the 
99\% level. Using the confidence level of 99\%, the SE model is
rejected only for $u=12\%$ and larger. This suggests that the SE is an adequate
representation of the distribution of drawdowns of the Nasdaq composite index
for drawdowns of amplitudes less than about $10\%$ at the 99\% confidence level,
i.e., $99\%$ of total number of drawdowns is described correctly by the SE.
Larger drawdowns require a 
different model as there is a statistically significant upward curvature detecting
by our test. This
test confirms our proposed picture of a change of regime between ``normal''
drawdowns (about $99\%$ of the total data set) of amplitude less than about $10\%$ 
to ``anomalous'' large drawdowns dubbed here ``outliers'' corresponding to 
about $1\%$ of the total data set.

The situation is more murky for the SP500 index as shown in table \ref{tablesp500}.
Here, already for the threshold $u=3\%$, corresponding to
$90\%$ of the total number of drawdowns, we see that the SE model is marginally 
rejected (or accepted) at the 99\% confidence level. For larger thresholds $u$,
the SE model is strongly rejected. The SE model has thus much less descriptive power
and the qualification of a change of regime is less clear.

For the Dow Jones Industrial Average (DJIA) index, the results
reported in table \ref{tabledjia} indicate that
the hypothesis $H_0$ that $C=0$ cannot be rejected at the $95\%$
confidence level for a cut-off
$u=3\%$ corresponding to $87\%$ of the total number of drawdowns. For
$u=6\%$, i.e., $97\%$ of the total data set,
the SE model is rejected at the 99\% confidence level.
This suggests that the SE is an adequate
representation of the distribution of drawdowns of the DJIA index
for drawdowns of amplitudes less than about $3-5\%$ at the 95\% confidence level,
i.e., about $90\%$ of total number of drawdowns is described correctly by the SE.
Larger drawdowns require a 
different model as there is a statistically significant upward curvature detecting
by our test. This
test again confirms our proposed picture of a change of regime between ``normal''
drawdowns (about $90\%$ of the total data set) of amplitude less than about $3-5\%$ 
to ``anomalous'' large drawdowns dubbed here ``outliers'' corresponding to 
about $10\%$ of the total data set. 

The conclusion that can be drawn from this analysis is that there is some support
for our proposal that large drawdowns are `outliers' but the issue is
made murky by the fact noticed also by Feigenbaum that the exponential or
stretched exponential model does not seem always to be a sufficiently good
model for the bulk
of the distribution of drawdowns. We stress however that the exponential distribution 
is the natural null hypothesis for uncorrelated returns, as shown in 
[Johansen et al., 2000b; Johansen and Sornette, 2001b]. This actually seems to be born
out approximately by our maximum likelihood estimates of the exponent $z$ of the 
stretched exponential model that determines it close to $1$ both for 
the S\&P500 (see table \ref{tablesp500}) and the DJIA (see table \ref{tabledjia})
while $z \approx 0.9$ for the Nasdaq index (see table \ref{tablenasdaq}). 
However, the fact that that the exponential distribution is only expected
asymptotically and the presence of dependences in the returns make the 
detection of outliers less clear-cut.

Therefore, any statistical test such as the one
we have presented on the possible existence of a change of regime in the 
distribution of drawdowns is 
by construction a joint test of both
the adequacy of the SE model and of the change of regime. This is 
unfortunate because one may be led to reject the hypothesis of a change of regime
while actually it is the model of the bulk of the distribution which is rejected.
To circumvent this problem, we need non-parametric tests of a change of regime
which are not dependent on the specific choice such as the stretched exponential
model used here. We will report on this approach in a future communication.

\section{Conclusion}

In addition to offering a synthesis of all our past results, excluding crashes
on emergent markets (see Johansen and Sornette [2001a]), the main 
novel achievements of this paper include:
\begin{enumerate}
\item nine new crash cases and a presentation of the spectrum of logperiodicity
 on 20 major cases;

\item a general solution of the rational expectation model with arbitrary risk
aversion quantified by an arbitrary stochastic pricing kernel;

\item a report on systematic out-of-sample tests over 20 years of data on the Hong
Kong index;

\item a precise mathematical assessment using probability theory of the statistical
signicance of a run of predictions using the ``crash roulette'';

\item a novel maximum likelihood test of our previous proposal that large drawdowns are
outliers.
\end{enumerate}

As this paper has shown, there are many subtle issues associated with 1) the concept
that crashes are critical events and 2) that the dynamics of stock market prices
develop specific log-periodic self-similar patterns (see [Ide and Sornette, 2001]
and [Gluzman and Sornette,2001] for recent theoretical developments). Nothwithstanding
6 years of work and tens of papers, the problem is still in its infancy and much
remains to be done to understand and use these critical log-periodic patterns.

\vskip 0.5cm
{\bf Acknowledgments}: We are grateful to V. F. Pisarenko for his advice on the
statistical testing procedure, to D. Farmer and
D. Stauffer for their encouragements and to W.-X. Zhou for
discussions. Notwithstanding
the firm tone adopted in our rebuttal to Feigenbaum's criticisms, we
acknowledge his efforts as a serious independent
statistical analysis of the log-periodic hypothesis.
This work was partially supported by the James S. Mc Donnell Foundation 21st century 
scientist award/studying complex system.

\newpage

\section*{Appendix A: GARCH(1,1)}

The GARCH(1,1) model of a price time series $x_{t}$ is defined as follows. 
\begin{equation}
x_{t}=\mu +\epsilon _{t}  \label{eqxt}
\end{equation}%
\begin{equation}
\epsilon _{t}=\sqrt{h_{t}}z_{t}\newline
\end{equation}%
\begin{equation}
h_{t}=\alpha +\beta \epsilon _{t-1}^{2}+\gamma h_{t-1}  \label{eqht}
\end{equation}%
where $z_{t}$ is drawn from the Student distribution with $\kappa =4$ degrees of freedom 
and mean $0$ and unit variance. Since 
the probability density function of the Student-t distribution is%
\begin{equation}
t\left( x,\kappa \right) =\frac{\Gamma \left( \frac{\kappa +1}{2}\right) }{%
\Gamma \left( \frac{\kappa }{2}\right) \sqrt{\kappa \pi }}\left( 1+\frac{%
x^{2}}{\kappa }\right) ^{-\frac{\kappa +1}{2}},
\end{equation}
the likelihood function $L$ of the scaled residual $z_{t}$ reads: 
\begin{eqnarray}
\frac{\log L}{T} &=&\log \left[ \Gamma \left( \frac{\kappa +1}{2}\right) %
\right] -\log \left[ \Gamma \left( \frac{\kappa }{2}\right) \right] -\frac{1%
}{2}\log (\kappa \pi )  \nonumber \\
&&{}-\frac{\kappa +1}{2T}\sum_{t=1}^{T}\left[ \log \left( 1+\frac{\epsilon
_{t}^{2}}{\kappa h_{t}}\right) \right] 
\end{eqnarray}%
where $\Gamma $ denotes the gamma function. The parameters of the 
GARCH(1,1) model are estimated using the maximum likelihood method. With
these parameters, the distribution of returns is well fitted by the GARCH model.
As shown in figure \ref{trailgarchconfcr}, the small and medium-size 
drawdowns are also reasonably accounted for by the GARCH(1,1) model.

\pagebreak

\section*{Appendix B: The crash roulette problem}

The ``crash roulette'' problem
defined in section \ref{crashroul}
is the same as the following game explained the book of W. Feller (1971).
In a population of $N$ 
balls, $n_c$ are red and $N-n_c$ are black. A group of $r$
balls is chosen at random. What is the probability $p_k$
that the group so chosen will contain exactly $k$ red balls?

Denoting $C(n,m) = n!/(m! (n-m)!$, the number of 
distinct ways to choose $m$ elements among $n$ elements, 
independently of the order with which we choose the $m$ elements,
we estimate the probability $p_k$.
If, among the $r$ chosen balls, there are $k$ red ones, then there
are $r-k$ black ones. There are thus $C(n_c,k)$ different ways
of choosing the red balls and $C(N-n_c,r-k)$ different ways of choosing
the black balls. The total number of ways of choosing $r$ balls
among $N$ is $C(N,r)$. Therefore, the probability $p_k$
that the group of $r$ balls so chosen will contain exactly $k$ red balls
is the product $C(n_c,k) \times C(N-n_c,r-k)$ of the number
ways corresponding
to the draw of exactly $k$ red balls among $r$ divided by the total possible number 
$C(N,r)$ of ways to draw the $r$ ball (here we simply use the
frequentist definition of
the probability of an event as the ratio of the number of states
corresponding to that event divided by the total number of events):
\be
p_k = {C(n_c,k) \times C(N-n_c,r-k) \over C(N,r)}~.
\ee
$p_k$ is the so-called Hypergeometric function.
In order to quantify a statistical confidence, we must ask a slightly
different question:
what is the probability $P_k$ that, out of the $r$ balls, there are
at least $k$ red balls? 
Clearly, the result is obtained by summing $p_k$ over
all possible values of $k$'s up to the maximum of $n_c$ and $r$;
indeed, the number of red balls among $r$ cannot be greater than 
$r$ and it cannot be greater than the total number $n_c$ of available red
balls.

\newpage

{\bf References}

\vskip 0.5cm

Adam, M.C. and A. Szafarz 1992 Speculative Bubbles and Financial Markets,
Oxford Economic Papers {\bf 44}, 626-640

Anifrani, J.-C., Le Floc'h, C., Sornette, D. and Souillard, B. 1995
Universal Log-periodic correction to renormalization group scaling 
for rupture stress
prediction from acoustic emissions, J.Phys.I France {\bf 5}, 631-638.

Anifrani, J.-C., Le Floc'h, C., Sornette, D. 1999
Pr\'ediction de la rupture de r\'eservoirs composites de haute 
pression \`a l'aide
de l'\'emission acoustique, Contr\^ole Industriel {\bf 220}, 43-45.

Blanchard, O.J. 1979 Speculative Bubbles, Crashes and Rational Expectations,
Economics Letters {\bf 3}, 387-389.

Blanchard, O.J. and M.W. Watson 1982 Bubbles, Rational Expectations and
Speculative Markets, in: Wachtel, P. ,eds., Crisis in Economic and Financial
Structure: Bubbles, Bursts, and Shocks. Lexington Books: Lexington.

Budyko, M.I. 1969 The effect of solar radiation variations
on the climate of the earth, Tellus {\bf 21}, 611-619.

Camerer, C. 1989 Bubbles and Fads in Asset Prices,
Journal of Economic Surveys {\bf 3}, 3-41

Canessa, E. 2000, Stochastic theory of log-periodic patterns,
J. Phys. A {\bf 33}, 9131-9140.
 
Cochrane, J.H. 2001, Asset Pricing (Princeton University Press, Princeton)

Drozdz, S., Ruf, F., Speth, J. and Wojcik, M. 1999
Imprints of log-periodic self-similarity in the stock market,
European Physical Journal {\bf 10}, 589-593.

Dubrulle, B., F. Graner and D. Sornette, eds. 1997
Scale invariance and beyond (EDP Sciences and Springer, Berlin).

Evans, G.W. 1991 Pitfalls in Testing for Explosive Bubbles
in Asset Prices, American Economic Review {\bf 81}, 922-930.

Feigenbaum, J.A. and Freund P.G.O. 1996
Discrete scale invariance in stock markets before crashes,
Int. J. Mod. Phys. B {\bf 10}, 3737-3745.

Feigenbaum J.A. and Freund P.G.O. 1998
Discrete scale invariance and the ''second black Monday'',
Modern Physics Letters B {\bf 12}, 57-60.

Feigenbaum, J.A. 2001 A statistical analysis of log-periodic precursors to
financial crashes, Quantitative Finance

Feller, W. 1971 {\it An Introduction to Probability Theory and
its Applications}. vol. I (John Wiley and sons,~New York), section 6 of chapter 2.

Garcimartin, A.,  Guarino, A.,  Bellon, L. and Ciliberto, S. 1997
Statistical properties of fracture precursors,
Phys. Rev. Lett. {\bf 79}, 3202-3205.

Ghil, M. and S. Childress 1987 Topics in 
Geophysical Fluid Dynamics:
Atmospheric Dynamics, Dynamo Theory and Climate Dynamics (Springer-Verlag, 
New York, Berlin, London, Paris, Tokyo).

Gluzman, S. and Yukalov, V.I. 1998
Booms and crashes in self-similar markets, Modern Physics Letters B 
{\bf 12}, 575-587.

Gluzman, S. and Sornette, D. 2001
Log-periodic route to fractal functions, submitted to Phys. Rev. E
(e-print at http://arXiv.org/abs/cond-mat/0106316)

Gopikrishnan, P., Meyer, M., Amaral, L.A.N. \& Stanley, H.E. 1998
Inverse Cubic Law for the Distribution of  Stock Price Variations,
European Physical Journal B 3, 139-140.

Guarino, A., Garcimartin, A. and Ciliberto, S. 1998
An experimental test of the critical behaviour of fracture precursors
Eur. Phys. J. B {\bf 6}, 13-24.

Guarino, A., Ciliberto, S. and Garcimartin, A. 1999
Failure time and microcrack nucleation,
Europhysics Letters {\bf 47}, 456-461.

Huang, Y., A. Johansen, M.W. Lee, H. Saleur and D. Sornette 2000a
Artifactual Log-Periodicity in Finite-Size Data: Relevance for 
Earthquake Aftershocks,
J. Geophys. Res. 105, 25451-25471.

Huang, Y., H. Saleur and D. Sornette 2000b
Reexamination of log-periodicity observed in the seismic precursors of the
1989 Loma Prieta earthquake, J. Geophysical Research 105, B12, 28111-28123.

Ide, K. and D. Sornette 2001
Oscillatory Finite-Time Singularities in Finance, Population and Rupture, 
preprint  (e-print at http://arXiv.org/abs/cond-mat/0106047)

Ilinski, K. 1999 Critical crashes? Int. J. Mod. Phys. C {\bf 10} 741-746.

Johansen,~A.,~Sornette,~D.,~Wakita,~G.,~Tsunogai,~U.,~Newman,~W.I. 
and Saleur,~H. 1996
Discrete scaling in earthquake precursory phenomena: evidence in the Kobe
earthquake,~Japan, J. Phys. I France {\bf 6},~1391-1402.

Johansen A. 1997  Discrete scale
invariance and other cooperative phenomena in spatially extended systems
with threshold dynamics,  Ph.D. Thesis, Niels Bohr Institute, University of
Copenhagen. Available from http://www.nbi.dk/\~\/johansen/pub.html.

Johansen, A. and D. Sornette, 1997
The footnote 12 of Stauffer D,
Sornette D. 1998. Log-periodic Oscillations for Biased Diffusion on 3D Random
Lattice, Physica A {\bf 252}, 271-277 reported the following: ``prediction of
the stock market turmoil at the end of October 1997, based on an unpublished
extension of the theory, have been formally issued ex-ante on September 17,
1997, to the French office for the  protection of proprietary softwares and
inventions under number registration 94781. In addition, a trading strategy
was been devised using put options in order to provide an experimental test of
the theory. A $400\%$ profit has been obtained in a two week period covering
the mini-crash of October 31, 1997. The proof of this profit is available from
a Merrill Lynch client cash management account released in November 1997. See
also Dupuis H.  {\em Un krach avant Novembre} in Tendances the 18. September
1997 page 26 refering to the work of Vandewalle et al. [1998b]
using the same type of log-periodic signals.
 
Johansen, A. and D. Sornette 1998a
Evidence of discrete scale invariance by canonical averaging,
Int. J. Mod. Phys. C {\bf 9}, 433-447.

Johansen, A. and Sornette, D. 1998b
Stock market crashes are outliers, European Physical Journal B {\bf 
1}, 141-143.

Johansen, A. and D. Sornette 1999a Critical Crashes, Risk 12 (1), 91-94

Johansen, A. and D. Sornette 1999b
Modeling the stock market prior to large crashes,
Eur. Phys. J. B {\bf 9}, 167-174.

Johansen A. and Sornette D. 1999c
Financial ``anti-bubbles'': log-periodicity in Gold and Nikkei collapses,
Int. J. Mod. Phys. C {\bf 10}, 563-575.

Johansen A., Sornette D. and  Ledoit O. 1999
Predicting Financial Crashes Using Discrete Scale Invariance, Journal 
of Risk {\bf 1}, 5-32.

Johansen, A. and D. Sornette 2000a
Critical ruptures, Eur. Phys. J. B {\bf 18}, 163-181.

Johansen, A. and Sornette, D. 2000b
The Nasdaq crash of April 2000: Yet another example of log-periodicity
in a speculative bubble ending in a crash,
  Euro. Phys. J. B {\bf 17}, 319-328.
 
Johansen A and D. Sornette D. 2000c
Evaluation of the quantitative prediction of a trend reversal on
the Japanese stock market in 1999, Int. J. Mod. Phys. C Vol. {\bf 11}, 359-364.
 
Johansen, A., H. Saleur and D. Sornette 2000a
New Evidence of Earthquake Precursory Phenomena in the
17 Jan. 1995 Kobe Earthquake, Japan, Eur. Phys. J. B 15, 551-555.

Johansen A, Ledoit O, Sornette D. 2000b Crashes as critical points,
International Journal of Theoretical and Applied Finance {\bf 3}, 219-255.

Johansen, A. and D. Sornette 2001a
Log-periodic power law bubbles in Latin-American and Asian markets
and correlated anti-bubbles in Western stock markets: An empirical study,
in press in Int. J. Theor. Appl. Fin.,
e-print at $http://xxx.lanl.gov/abs/cond-mat/9907286$

Johansen, A. and D. Sornette 2001b
Large Stock Market Price Drawdowns Are Outliers, in press in the
Journal of Risk
(preprint at $http://arXiv.org/abs/cond-mat/0010050$).

L'vov, V.S., Pomyalov, A. and Procaccia, I. 2001
Outliers, Extreme Events and Multiscaling, Phys. Rev. E 6305, 
PT2:6118, U158-U166 (2001).

Laloux, L., Potters, M., Cont, R., Aguilar, J.P. and Bouchaud, J.-P. 1999
Are financial crashes predictable? Europhys. Lett. {\bf 45}, 1-5.

Lillo, F., and Mantegna, R.N. 2000 
Symmetry alteration of ensemble return distribution in crash and rally
days of financial markets, European Physical Journal B {\bf 15}, 603-606.

Lux, T. and D. Sornette 2001
On Rational Bubbles and Fat Tails, in press in the Journal
of Money, Credit and Banking
(e-print at $http://xxx.lanl.gov/abs/cond-mat/9910141$)

Main, I.G., O'Brien, G. and Henderson, J.R. 2000
Statistical physics of earthquakes: Comparison of distribution exponents
for source area and potential energy and the dynamic emergence of
log-periodic energy quanta, J. Geophys. Res. {\bf 105}, 6105-6126.
 
Mandelbrot, B.B. 1963 The variation of certain speculative prices,
Journal of Business {\bf 36}, 392-417.

Mansilla, R., Algorithmic Complexity in Real Financial Markets,
cond-mat/0104472

Mantegna, R.N. and Stanley, H.E. 1995 Scaling behaviour in the dynamics
of an economic index, Nature {\bf 376}, 46-49.

Molchan, G.M. (1990) Strategies in strong earthquake
prediction, Physics of the Earth and Planetary Interiors {\bf 61}, 84-98 (1990).

Molchan, G.M. (1997) Earthquake prediction as a
decision-making problem, Pure and Applied Geophysics {\bf 149}, 233-247 (1997).

Newman, W.I., D.L. Turcotte and A.M. Gabrielov 1995
Log-periodic behavior of a hierarchical failure model with applications to
precursory seismic activation, Phys. Rev. E {\bf 52}, 4827-4835.

Press, W.H. {\it et al.~}, Numerical Recipes Cambridge University Press (1992).

Rao, C. 1965 Linear statistical inference and its 
applications, Wiley, N.Y., 522 p. (Chapter 6, Section 6e.3).

Sahimi, M. and S. Arbabi 1996 Scaling laws for fracture of
heterogeneous materials and rocks,
Phys. Rev. Lett. {\bf 77}, 3689-3692.

Saleur,~H.,~Sammis,~C.G. and Sornette,~D. (1996)
Discrete scale invariance,~complex fractal dimensions and log-periodic
corrections in earthquakes, J. Geophys. Res. {\bf 101},~17661-17677.
 
Sellers, W.D. 1969 A climate model based on the energy
balance of the earth-atmosphere system, J. Appl. Meteorol. {\bf 8}, 392-400.

Sobolev G.A. and Yu. S. Tyupkin 2000 Analysis of energy release process during
main rupture formation in laboratory studies of rock fracture and before
strong earthquakes, Izvestiya, Physics of the Solid Earth {\bf 36}, 138-149.

Sornette,~D. and Sammis,~C.G. 1995
Complex critical exponents from renormalization group theory of
earthquakes: Implications for earthquake predictions, J. Phys. I 
France {\bf 5},
607-619.

Sornette, D., A. Johansen and J.-P. Bouchaud 1996
Stock market crashes, Precursors and Replicas, J.Phys.I France {\bf 
6}, 167-175.
 
Sornette D. and Johansen A. 1997
Large financial crashes, Physica A {\bf 245}, 411-422.

Sornette D. 1998 Discrete scale invariance
and complex dimensions, Phys. Rep. {\bf 297}, 239-270.

Sornette D. and Johansen A. 1998
A Hierarchical Model of Financial Crashes, Physica A {\bf 261}, 581-598.

Sornette, D. and Y. Malevergne 2001
>From Rational Bubbles to Crashes, submitted to Physica A
(e-print at $http://arXiv.org/abs/cond-mat/0102305$)

Sornette, D. and K. Ide 2001
Theory of self-similar oscillatory finite-time singularities
in Finance, Population and Rupture, preprint
(e-print at http://arXiv.org/abs/cond-mat/0106054)

US Committee of the 
Global Atmospheric Research Program 1975 National Research Council,
Understanding climatic change -- A program for action, 
National Academy of Sciences, Washington, D.C.

Vandewalle, N., Boveroux, P., Minguet, A. and Ausloos, M. 1998a
The crash of October 1987 seen as a phase transition: amplitude and
universality, Physica A {\bf 255}, 201-210.

Vandewalle N, Ausloos M,  Boveroux Ph, Minguet A. 1998b
How the financial crash of October 1997 could have been
predicted. European Physics Journal B {\bf 4}: 139-141.

Vandewalle, N., Ausloos, M., Boveroux, P. and Minguet, A. 1999
Visualizing the log-periodic pattern before crashes,
European Physical Journal B {\bf 9}, 355-359.
 
Vries, C.G. de 1994 Stylized Facts of Nominal Exchange
Rate Returns, S. 348 - 89 in: van der Ploeg, F., ed.,
The Handbook of International Macroeconomics. Blackwell: Oxford.

Woo, W.T. 1987 Some evidence of speculative bubbles in the foreign
exchange markets, Journal of Money, Credit and Banking {\bf 19}, 499-514.

\pagebreak

\clearpage

\clearpage

\clearpage

\begin{table}[]
\begin{center}
\begin{tabular}{|c|c|c|c|c|c|c|} \hline
Cut-Off $u$  & quantile & $z$  & $\ln\lp L_0\rp$ & $\ln\lp L_1\rp$ & $T$ & proba \\ \hline
$3\%$ & $87\%$   & $0.916,0.940$ & $4890.36$ & $4891.16$ &  $1.6$ & $20.5\%$  \\ \hline
$6\%$ & $97\%$   & $0.875,0.915$ & $4944.36$ & $4947.06$ &  $5.4$ & $2.0\%$   \\ \hline
$9\%$ & $99.0\%$ & $0.869,0.918$ & $4900.75$ & $4903.66$ &  $5.8$ & $1.6\%$   \\ \hline
$12\%$& $99.7\%$ & $0.851,0.904$ & $4872.47$ & $4877.46$ &  $10.0$ & $0.16\%$ \\ \hline
$15\%$& $99.7\%$ & $0.843,0.898$ & $4854.97$ & $4860.77$ &  $11.6$ & $0.07\%$ \\ \hline
$18\%$& $99.9\%$ & $0.836,0.890$ & $4845.16$ & $4851.94$ &  $13.6$ & $0.02\%$ \\ \hline
\end{tabular}
\vspace{5mm}
\caption{\label{tablenasdaq} Nasdaq composite index. The total number af
drawdowns is $1495$. The first column is the cut-off
$u$ such that the MLE of the two competing hypotheses (standard (SE) and modified (MSE)
stretched exponentials) is performed over the interval $\left[0, u\right]$ of the absolute
value of the drawdowns. The second column gives the fraction `quantile' of the
drawdowns belonging to $\left[0, u\right]$. The third column gives the exponents $z$ found
for the SE (first value) and MSE (second value) distributions. The fourth and fifth
columns give the
logarithm of the likelihoods (\ref{aaa}) and (\ref{bbb}) for the SE and MSE,
respectively. The sixth
column gives the variable $T$ defined in (\ref{testat}). The last column `proba' gives the 
corresponding probability of exceeding $T$ by chance. For $u>18\%$, we find that
$T$ saturates to $13.6$  and `proba' to $0.02\%$. 
}
\end{center}
\end{table}

\begin{table}[]
\begin{center}
\begin{tabular}{|c|c|c|c|c|c|c|} \hline
Cut-Off $u$  & quantile & $z$ & $\ln\lp L_0\rp$ & $\ln\lp L_1\rp$ & $T$ & proba \\ \hline
$3\%$ & $90\%$   & $1.05,1.08$ & $11438.55$ & $11442.11$ &  $7.1$ & $0.8\%$ \\ \hline
$6\%$ & $98.6\%$ & $0.981,1.04$& $11502.00$ & $11511.95$ &  $19.9$ & $<10^{-4}\%$\\ \hline
$9\%$ & $99.6\%$ & $0.971,1.03$& $11441.17$ & $11451.72$ &  $21.1$ & $<10^{-4}\%$\\ \hline
$12\%$& $99.9\%$ & $0.960,1.02$& $11405.89$ & $11417.62$ &  $23.5$ & $<10^{-4}\%$\\ \hline
$15\%$& $99.97\%$& $0.956,1.01$& $11394.67$ & $11407.67$ &  $26.0$ & $<10^{-4}\%$\\ \hline
\end{tabular}
\vspace{5mm}
\caption{\label{tablesp500} SP500 index. Same as table \protect\ref{tablenasdaq}.
The total number af drawdowns is $3363$. For $u>15\%$, we find that $T$ saturates to 
$26$  and `proba' is less than $10^{-6}$.
}
\end{center}
\end{table}

\begin{table}[]
\begin{center}
\begin{tabular}{|c|c|c|c|c|c|c|} \hline
Cut-Off $u$  & quantile & $z$ & $\ln\lp L_0\rp$ & $\ln\lp L_1\rp$ & $T$ & proba \\ \hline
$3\%$ & $87\%$   & $1.01,1.04$  & $21166.73$ & $21168.46$ &  $3.5$ & $6.1 \%$ \\ \hline
$6\%$ & $97\%$   & $0.965,1.01$ & $21415.28$ & $21424.00$ &  $17.4$ & $0.003\%$ \\ \hline
$9\%$ & $99.3\%$ & $0.934,0.992$& $21229.22$ & $21248.65$ &  $38.9$ & $<10^{-4}\%$ \\ \hline
$12\%$& $99.8\%$ & $0.921,0.982$& $21132.75$ & $21157.04$ &  $48.6$ & $<10^{-4}\%$ \\ \hline
$15\%$& $99.9\%$ & $0.917,0.977$& $21100.87$ & $21126.13$ &  $50.5$ & $<10^{-4}\%$ \\ \hline
$18\%$& $99.9\%$ & $0.915,0.973$& $21089.15$ & $21114.08$ &  $49.9$ & $<10^{-4}\%$ \\ \hline
$21\%$& $99.95\%$& $0.912,0.966$& $21075.20$ & $21100.88$ &  $51.5$ & $<10^{-4}\%$ \\ \hline
$24\%$& $99.97\%$& $0.910,0.957$& $21065.62$ & $21090.57$ &  $49.9$ & $<10^{-4}\%$ \\ \hline
\end{tabular}
\vspace{5mm}
\caption{\label{tabledjia} DJIA index. Same as tables \protect\ref{tablenasdaq} and 
\protect\ref{tablesp500}. The total number of drawdowns is $6486$. For $u>24\%$, we 
find that $T$ saturates to $50$  and `proba' is less than $10^{6}$.
}
\end{center}
\end{table}   

\end{document}